\documentclass[hyper]{JHEP3}
\pdfoutput=1
\usepackage{amsmath,amssymb,amsfonts,amsxtra,mathrsfs,makeidx,graphics,graphicx,amsthm,bbm}
\usepackage{youngtab}

\def\one{ 
\setlength{\unitlength}{0.45cm} 
\begin{picture}(0.55,0.5) 
\put(0,0){\line(1,0){0.4}} 
\put(0,.4){\line(1,0){0.4}} 
\multiput(0,0)(.4,0){2}{\line(0,1){.4}} 
\end{picture}} 
\def\twohor{ 
\setlength{\unitlength}{0.45cm} 
\begin{picture}(.9,0.5) 
\put(0,0){\line(1,0){0.8}} 
\put(0,.4){\line(1,0){0.8}} 
\multiput(0,0)(.4,0){3}{\line(0,1){.4}} 
\end{picture}} 
\def\twover{ 
\setlength{\unitlength}{0.45cm} 
\begin{picture}(0.55,0.5) 
\put(0,0){\line(1,0){0.4}} 
\put(0,.4){\line(1,0){0.4}} 
\put(0,-.4){\line(1,0){0.4}} 
\multiput(0,0)(.4,0){2}{\line(0,0){.4}} 
\multiput(0,0)(.4,0){2}{\line(0,-1){.4}} 
\end{picture}}

\preprint{}

\title{Reformulated invariants for non-torus knots \& links}

\author{Zodinmawia and P. Ramadevi\\
Department of Physics, Indian Institute of Technology Bombay,\\
 Mumbai, India, 400076\\ \\
 {\tt\email{zodin@phy.iitb.ac.in, ramadevi@phy.iitb.ac.in}}}

\abstract{Using the Racah coefficients in our earlier paper \href{http://arxiv.org/abs/1107.3918}{arXiv:11073918},  we explicitly write the  
Chern-Simons field theory invariants for many non-torus 
knot and links. Further, we have tabulated  the reformulated invariants which agrees with the
Ooguri-Vafa conjecture  for knots
 and  Labastida-Marino-Vafa conjecture for links. 
}


\makeatother

\begin{document}

\maketitle

\section{Introduction}
\label{sec:1}
Starting with the pioneering work of Witten \cite{Witten:1988hf}, there have interesting developments between
Chern-Simons theory and knot theory.
$U(N)$ Chern-Simons theory provides a pool of colored framed knot invariants $V_R[\mathcal{K};p]$  given by the expectation value of
Wilson loop operator $\mathcal{K}$, with framing number $p$, carrying representation $R$ \cite{Borhade:2003cu}. We can place  different representations 
 on the components  of a $r$-component link  giving multicolored framed  link invariants $V_{R_1,R_2 ,\ldots R_r}[\mathcal{L};p_1,p_2,\ldots p_r]$
where the $r$-tuple vector $\vec p$ denotes the framing numbers on the respective  component knots. 
In this paper, we will write the invariants for zero framed knots (p=0) and zero framed 2-component links ($p_1=p_2=0$).

The polynomial form of these colored invariants can be obtained for  $(2,2p+1)$ class of torus knots
and $(2,2p)$ class of torus links \cite{RamaDevi:1992dh,Ramadevi:1996:PHD,Ramadevi:2000gq,chandrima,Paul:2010wr}. 
For the most general class of $(n,m)$ torus knots , the explicit polynomial 
form for  fundamental representation  \Yboxdim5pt($R=\yng(1)$)   placed on the knot is given  by Theorem  9.7 in Ref.\cite{Jones:1987dy} and from Chern-Simons theory \cite{nmtorus5}.
Further,  the colored invariants of $(n,m)$ torus knots and links have been discussed in  Refs.\cite{Labastida:2000zp,nmtorus1,nmtorus2,nmtorus3,nmtorus4}. 

\begin{figure}
\begin{centering}
\includegraphics{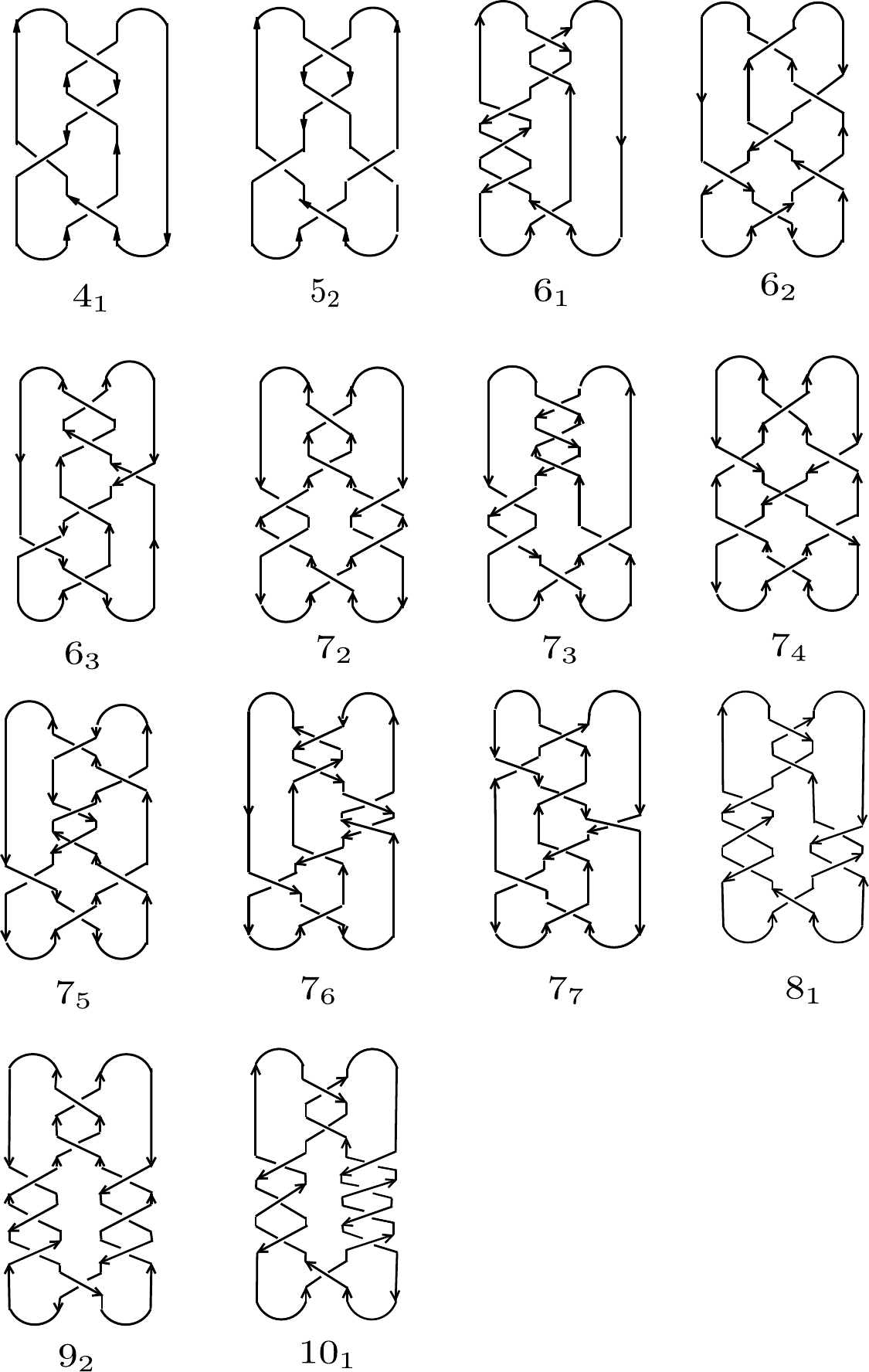}\caption{Non-torus knots.\label{fig:knots}}
\par\end{centering}
\end{figure}

\begin{figure}
\centering{}\includegraphics{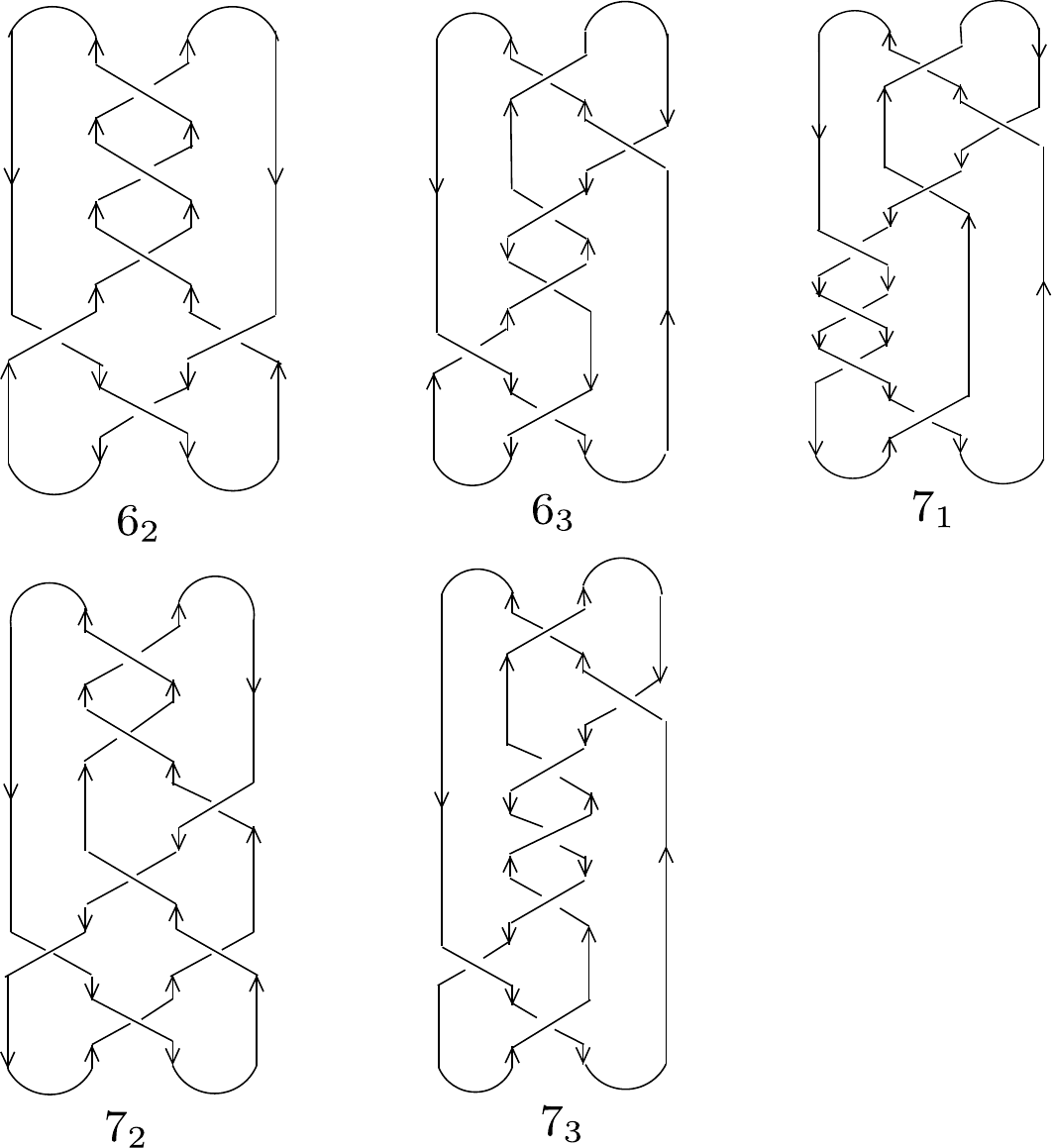}\caption{Non-torus links.\label{fig:links}}
\end{figure}
For non-torus  knots and non-torus links, we could formally write an expression for these  invariants which involves
quantum Racah coefficients \cite{RamaDevi:1992dh,Ramadevi:1996:PHD}.  Plat  diagrams  for various non-torus knots and
links  are  given in Figure \ref{fig:knots}
and Figure \ref{fig:links}. Recently, the colored  invariants of figure eight ($\bf{4_1}$) knot carrying totally symmetric or totally antisymmetric representation
has been conjectured \cite{Itoyama:2012fq}. However, for other non-torus knots we could write 
the explicit polynomial form only for some representations using the  Racah coefficients \cite{Zodinmawia:2011ud}.
We present, for the knots and links in Figures \ref{fig:knots}, \ref{fig:links}, 
 the polynomial form of the colored knot invariants and multicolored two-component  link invariants.

Using these colored knot and multicolored link  invariants, we tabulate the reformulated invariants. The polynomial
form of these reformulated invariants  indeed agrees with  the Ooguri-Vafa conjecture for knots \cite{Ooguri:1999bv} 
and Labastida-Marino-Vafa for links \cite{Labastida:2000yw}.
The results indirectly confirms that the Racah coefficients obtained in \cite{Zodinmawia:2011ud} from the equivalence of knots or links are correct. 

The plan of the paper is as follows: In section \ref{sec:2}, we present the Chern-Simons invariants of the non-torus knots and links in Figures \ref{fig:knots}, \ref{fig:links}.
Then in section \ref{sec:3}, we write down the polynomial form of these invariants for some representations. Using the polynomial form, we 
obtain  the reformulated polynomials in section \ref{sec:4}. We find the reformulated polynomials of knots obey Ooguri-Vafa conjecture
and that of links obey Labastida-Marino-Vafa conjecture. Finally, we present  some of the challenging open problems in the concluding section.  

\section{U(N) invariants in terms of the quantum  Racah coefficients}
\label{sec:2}
$U(N)$ Chern-Simons invariant,  for any knot or link,   is a product of  $SU(N)$
and the $U(1)$ invariant \cite{Borhade:2003cu}. The $U(1)$ link invariant is related
to the linking and self linking number of the component knots. With a suitable choice of $U(1)$ charge \cite{Marino:2001re}, the $U(N)$  invariants are polynomials in variable $q=e^{\frac{2{\pi}i}{k+N}}$ and $\lambda=q^N$ where $k$ is the Chern-Simons coupling.
Following  Ref.\cite{RamaDevi:1992dh}, the  $SU(N)$ link invariant
can be directly written down using the two inputs:(1)
The relation between $SU(N)$ Chern Simons field theory on a three,
ball with $SU(N)_{k}$ conformal field theory on the boundary of the
three-ball \cite{Witten:1988hf}, (2) any 
knot or and link can be obtained as platting of braids (Birman theorem) \cite{birman}. 
Explicit discussion of this method can be found in \cite{Zodinmawia:2011ud} where we have 
obtained the $U(N)$  invariant  for the framed knot $5_2$ and the  framed link $6_2$.
Particularly, the braiding eigenvalues of parallely oriented strands $\lambda_s^{(+)}(R_1,R_2)$ and
antiparallely oriented strands $\lambda_s^{(-)}(R_1,\bar {R_2})$ (see equation (2.14) in \cite{Zodinmawia:2011ud}) and the
Racah coefficients  presented \cite{Zodinmawia:2011ud} will be useful to write the 
polynomial form for some representations. 

For completeness, we present  the expression of $U(N)$ link invariant for
non torus knots  and links in Figure \ref{fig:knots} and Figure \ref{fig:links}
in terms of $SU(N)$ quantum Racah coefficients and braiding eigenvalues. 
\subsection{Non Torus Knots}
In these formulae, $l$ denotes the total number of boxes in the Young Tableaux representation $R$  and
$\kappa_R= \frac{1}{2}[Nl+l+\sum_i(l_i^2-2il_i)]$  where $l_i$ is the number of boxes in the i-th row. Using the braiding eigenvalues,
Racah coefficients, $\kappa_R$ and $l$, we write the $U(N)$ Chern-Simons  invariant for zero-framed non-torus knots:
\begin{flalign}
V_{R}^{\{U(N)\}}[4_{1};\,0] = & \sum_{s,t,s^{\prime}}\epsilon_{s}^{\bar{R},R}\,\sqrt{dim_{q}s}\,\epsilon_{s^{\prime}}^{R,R}\,\sqrt{dim_{q}s^{\prime}}\,(\lambda_{s}^{(-)}(\bar{R},\, R))^{2}\, a_{ts}\left[\begin{array}{cc}
R & \bar{R}\\
R & \bar{R}
\end{array}\right]&\nonumber\\
   &(\lambda_{t}^{(-)}(\bar{R},\, R))^{-1}\, a_{ts^{\prime}}\left[\begin{array}{cc}
\bar{R} & R\\
R & \bar{R}
\end{array}\right]\,(\lambda_{s^{\prime}}^{(+)}(R,\, R))^{-1}.&
\end{flalign}
\begin{flalign}
V_{R}^{\{U(N)\}}[5_{2};\,0] = & q^{(-5\kappa_{R}+\frac{5l^{2}}{2N})}\sum_{s,t,s^{\prime}}\epsilon_{s}^{R,R}\,\sqrt{dim_{q}s}\,\epsilon_{s^{\prime}}^{R,R}\,\sqrt{dim_{q}s^{\prime}}\,(\lambda_{s}^{(+)}(R,\, R))^{-2}\, a_{ts}\left[\begin{array}{cc}
\bar{R} & R\\
R & \bar{R}
\end{array}\right]&\nonumber \\
   & (\lambda_{t}^{(-)}(\bar{R},\, R))^{-2}\, a_{ts^{\prime}}\left[\begin{array}{cc}
R & \bar{R}\\
\bar{R} & R
\end{array}\right]\,(\lambda_{s^{\prime}}^{(+)}(R,\, R))^{-1}.&
\end{flalign}
\begin{flalign}
V_{R}^{\{U(N)\}}[6{}_{1};\,0] = & q^{(-2\kappa_{R}+\frac{l^{2}}{N})}\sum_{s,t,s^{\prime}}\epsilon_{s}^{\bar{R},R}\,\sqrt{dim_{q}s}\,\epsilon_{s^{\prime}}^{R,R}\,\sqrt{dim_{q}s^{\prime}}\,(\lambda_{s}^{(-)}(\bar{R},\, R))^{2}\, a_{ts}\left[\begin{array}{cc}
R & \bar{R}\\
R & \bar{R}
\end{array}\right]&\nonumber \\
   & (\lambda_{t}^{(-)}(\bar{R},\, R))^{-3}\, a_{ts^{\prime}}\left[\begin{array}{cc}
\bar{R} & R\\
R & \bar{R}
\end{array}\right]\,(\lambda_{s^{\prime}}^{(+)}(R,\, R))^{-1}.&
\end{flalign}
\begin{flalign}
V_{R}^{\{U(N)\}}[6_{2};\,0] = & q^{(-2\kappa_{R}+\frac{l^{2}}{N})}\sum_{s,t,s^{\prime},u,v}\epsilon_{s}^{R,R}\,\sqrt{dim_{q}s}\,\epsilon_{v}^{\bar{R},R}\,\sqrt{dim_{q}v}\,\lambda_{s}^{(+)}(R,\, R)\, a_{ts}\left[\begin{array}{cc}
\bar{R} & R\\
R & \bar{R}
\end{array}\right]&\nonumber \\
   & \lambda_{t}^{(-)}(R,\,\bar{R})\, a_{ts^{\prime}}\left[\begin{array}{cc}
\bar{R} & R\\
\bar{R} & R
\end{array}\right]\,(\lambda_{s^{\prime}}^{(-)}(R,\,\bar{R}))^{-1}\, a_{us^{\prime}}\left[\begin{array}{cc}
\bar{R} & \bar{R}\\
R & R
\end{array}\right]&\nonumber \\
   & (\lambda_{u}^{(+)}(R,\, R))^{-2}a_{uv}\left[\begin{array}{cc}
\bar{R} & \bar{R}\\
R & R
\end{array}\right]\,(\lambda_{v}^{(-)}(\bar{R},\, R))^{-1}.&
\end{flalign}
\begin{flalign}
V_{R}^{\{U(N)\}}[6_{3};\,0] = & \sum_{s,t,s^{\prime},u,v}\epsilon_{s}^{R,R}\,\sqrt{dim_{q}s}\,\epsilon_{v}^{\bar{R},R}\,\sqrt{dim_{q}v}\,(\lambda_{s}^{(+)}(R,\, R))^{-2}\, a_{ts}\left[\begin{array}{cc}
\bar{R} & R\\
R & \bar{R}
\end{array}\right]&\nonumber \\
   & (\lambda_{t}^{(-)}(R,\,\bar{R}))^{-1}\, a_{ts^{\prime}}\left[\begin{array}{cc}
\bar{R} & R\\
\bar{R} & R
\end{array}\right]\,\lambda_{s^{\prime}}^{(-)}(R,\,\bar{R})\, a_{us^{\prime}}\left[\begin{array}{cc}
\bar{R} & \bar{R}\\
R & R
\end{array}\right]&\nonumber \\
   & (\lambda_{u}^{(+)}(\bar{R},\,\bar{R}))a_{uv}\left[\begin{array}{cc}
\bar{R} & \bar{R}\\
R & R
\end{array}\right]\,(\lambda_{v}^{(-)}(\bar{R},\, R)).&
\end{flalign}
\begin{flalign}
V_{R}^{\{U(N)\}}[7_{2};\,0]=& q^{(-7\kappa_{R}+\frac{7l^{2}}{2N})}\sum_{s,t,s^{\prime}}\epsilon_{s}^{R,R}\,\sqrt{dim_{q}s}\,\epsilon_{s^{\prime}}^{R,R}\,\sqrt{dim_{q}s^{\prime}}\,(\lambda_{s}^{(+)}(R,\, R))^{-2}\, a_{ts}\left[\begin{array}{cc}
\bar{R} & R\\
R & \bar{R}
\end{array}\right]&\nonumber \\
  &(\lambda_{t}^{(-)}(R,\,\bar{R}))^{-4}\, a_{ts^{\prime}}\left[\begin{array}{cc}
\bar{R} & R\\
R & \bar{R}
\end{array}\right]\,(\lambda_{s^{\prime}}^{(+)}(R,\, R))^{-1}.&
\end{flalign}
\begin{flalign}
V_{R}^{\{U(N)\}}[7_{3};\,0] =&q^{(7\kappa_{R}-\frac{7l^{2}}{2N})} \sum_{s,t,s^{\prime}}\epsilon_{s}^{R,\bar{R}}\,\sqrt{dim_{q}s}\,\epsilon_{s^{\prime}}^{R,\bar{R}}\,\sqrt{dim_{q}s^{\prime}}\,(\lambda_{s}^{(-)}(R,\,\bar{R}))^{3}\, a_{ts}\left[\begin{array}{cc}
\bar{R} & \bar{R}\\
R & R
\end{array}\right]&\nonumber \\
   & (\lambda_{t}^{(+)}(\bar{R},\,\bar{R}))^{3}\, a_{ts^{\prime}}\left[\begin{array}{cc}
\bar{R} & \bar{R}\\
R & R
\end{array}\right]\,\lambda_{s^{\prime}}^{(-)}(R,\,\bar{R}).&
\end{flalign}
\begin{flalign}
V_{R}^{\{U(N)\}}[7_{4},;\,0] = & q^{(7\kappa_{R}-\frac{7l^{2}}{2N})}\sum_{s,t,s^{\prime},u,v}\epsilon_{s}^{R,R}\,\sqrt{dim_{q}s}\,\epsilon_{v}^{R,R}\,\sqrt{dim_{q}v}\,\lambda_{s}^{(+)}(R,\, R)\, a_{ts}\left[\begin{array}{cc}
\bar{R} & R\\
R & \bar{R}
\end{array}\right]&\nonumber \\
   & (\lambda_{t}^{(-)}(R,\,\bar{R}))^{2}\, a_{ts^{\prime}}\left[\begin{array}{cc}
R & \bar{R}\\
\bar{R} & R
\end{array}\right]\,\lambda_{s^{\prime}}^{(+)}(\bar{R},\,\bar{R})\, a_{us^{\prime}}\left[\begin{array}{cc}
R & \bar{R}\\
\bar{R} & R
\end{array}\right]&\nonumber \\
   & (\lambda_{u}^{(-)}(R,\,\bar{R}))^{2}\, a_{uv}\left[\begin{array}{cc}
\bar{R} & R\\
R & \bar{R}
\end{array}\right]\,\lambda_{v}^{(+)}(R,\, R).&
\end{flalign}
\begin{flalign}
V_{R}^{\{U(N)\}}[7_{5};\,0] =&q^{(-7\kappa_{R}+\frac{7l^{2}}{2N})}  \sum_{s,t,s^{\prime},u,v}\epsilon_{s}^{R,\bar{R}}\,\sqrt{dim_{q}s}\,\epsilon_{v}^{R,\bar{R}}\,\sqrt{dim_{q}v}\,(\lambda_{s}^{(-)}(R,\, R))^{-1}\, a_{ts}\left[\begin{array}{cc}
\bar{R} & \bar{R}\\
R & R
\end{array}\right]&\nonumber \\
   & (\lambda_{t}^{(+)}(R,\, R))^{-1}\, a_{ts^{\prime}}\left[\begin{array}{cc}
\bar{R} & \bar{R}\\
R & R
\end{array}\right]\,(\lambda_{s^{\prime}}^{(-)}(\bar{R},\, R))^{-2}\, a_{us^{\prime}}\left[\begin{array}{cc}
\bar{R} & \bar{R}\\
R & R
\end{array}\right]&\nonumber \\
   & \lambda_{u}^{(+)}(R,\, R)\, a_{uv}\left[\begin{array}{cc}
\bar{R} & \bar{R}\\
R & R
\end{array}\right]\,(\lambda_{v}^{(-)}(\bar{R},\, R))^{-1}.&
\end{flalign}
\begin{flalign}
V_{R}^{\{U(N)\}}[7_{6};\,0]= & q^{(-3\kappa_{R}+\frac{3l^{2}}{2N})}\sum_{s,t,s^{\prime},u,v}\epsilon_{s}^{R,\bar{R}}\,\sqrt{dim_{q}s}\,\epsilon_{v}^{\bar{R},R}\,\sqrt{dim_{q}v}\,(\lambda_{s}^{(-)}(R,\,\bar{R}))^{-2}\, a_{ts}\left[\begin{array}{cc}
\bar{R} & R\\
\bar{R} & R
\end{array}\right]&\nonumber \\
   & (\lambda_{t}^{(-)}(\bar{R},\, R))^{2}\, a_{ts^{\prime}}\left[\begin{array}{cc}
\bar{R} & R\\
\bar{R} & R
\end{array}\right]\,(\lambda_{s^{\prime}}^{(-)}(R,\,\bar{R}))^{-1}\, a_{us^{\prime}}\left[\begin{array}{cc}
\bar{R} & \bar{R}\\
R & R
\end{array}\right]&\nonumber \\
   & (\lambda_{u}^{(+)}(\bar{R},\,\bar{R}))^{-1}\, a_{uv}\left[\begin{array}{cc}
\bar{R} & \bar{R}\\
R & R
\end{array}\right]\,(\lambda_{v}^{(-)}(\bar{R},\, R))^{-1}.&
\end{flalign}
\begin{flalign}
V_{R}^{\{U(N)\}}[7_{7};\,0] = & q^{(\kappa_{R}-\frac{l^{2}}{2N})}\sum_{s,t,s^{\prime},u,v,w,x}\epsilon_{s}^{R,R}\,\sqrt{dim_{q}s}\,\epsilon_{v}^{R,R}\,\sqrt{dim_{q}v}\,(\lambda_{s}^{(+)}(R,\, R))\, a_{ts}\left[\begin{array}{cc}
\bar{R} & R\\
\bar{R} & R
\end{array}\right]&\nonumber \\
   & (\lambda_{t}^{(-)}(\bar{R},\, R))\, a_{ts^{\prime}}\left[\begin{array}{cc}
\bar{R} & R\\
R & \bar{R}
\end{array}\right]\,(\lambda_{s^{\prime}}^{(-)}(\bar{R},\, R))^{-1}\, a_{us^{\prime}}\left[\begin{array}{cc}
R & \bar{R}\\
R & \bar{R}
\end{array}\right]&\nonumber \\
   & (\lambda_{u}^{(+)}(\bar{R},\,\bar{R}))^{-1}\, a_{uv}\left[\begin{array}{cc}
R & R\\
\bar{R} & \bar{R}
\end{array}\right]\,(\lambda_{v}^{(-)}(R,\,\bar{R}))^{-1}a_{wv}\left[\begin{array}{cc}
R & \bar{R}\\
R & \bar{R}
\end{array}\right]&\nonumber \\
   & \lambda_{w}^{(-)}(R,\bar{R})\, a_{wx}\left[\begin{array}{cc}
\bar{R} & R\\
R & \bar{R}
\end{array}\right]\,\lambda_{x}^{(+)}(R,R).&
\end{flalign}
\begin{flalign}
V_{R}^{\{U(N)\}}[8{}_{1};\,0]= & q^{(-4\kappa_{R}+\frac{2l^{2}}{N})}\sum_{s,t,s^{\prime}}\epsilon_{s}^{\bar{R},R}\,\sqrt{dim_{q}s}\,\epsilon_{s^{\prime}}^{R,R}\,\sqrt{dim_{q}s^{\prime}}\,(\lambda_{s}^{(-)}(\bar{R},\, R))^{2}\, a_{ts}\left[\begin{array}{cc}
R & \bar{R}\\
R & \bar{R}
\end{array}\right]&\nonumber \\
   & (\lambda_{t}^{(-)}(\bar{R},\, R))^{-5}\, a_{ts^{\prime}}\left[\begin{array}{cc}
\bar{R} & R\\
R & \bar{R}
\end{array}\right]\,(\lambda_{s^{\prime}}^{(+)}(R,\, R))^{-1}.&
\end{flalign}
\begin{flalign}
V_{R}^{\{U(N)\}}[9_{2};\,0]= & q^{(-7\kappa_{R}+\frac{7l^{2}}{2N})}\sum_{s,t,s^{\prime}}\epsilon_{s}^{R,R}\,\sqrt{dim_{q}s}\,\epsilon_{s^{\prime}}^{R,R}\,\sqrt{dim_{q}s^{\prime}}\,(\lambda_{s}^{(+)}(R,\, R))^{-2}\, a_{ts}\left[\begin{array}{cc}
\bar{R} & R\\
R & \bar{R}
\end{array}\right]&\nonumber \\
   & (\lambda_{t}^{(-)}(\bar{R},\, R))^{-6}\, a_{ts^{\prime}}\left[\begin{array}{cc}
R & \bar{R}\\
\bar{R} & R
\end{array}\right]\,(\lambda_{s^{\prime}}^{(+)}(R,\, R))^{-1}.&
\end{flalign}
\begin{flalign}
V_{R}^{\{U(N)\}}[10{}_{1};\,0] = & q^{(-6\kappa_{R}+\frac{3l^{2}}{N})}\sum_{s,t,s^{\prime}}\epsilon_{s}^{\bar{R},R}\,\sqrt{dim_{q}s}\,\epsilon_{s^{\prime}}^{R,R}\,\sqrt{dim_{q}s^{\prime}}\,(\lambda_{s}^{(-)}(\bar{R},\, R))^{2}\, a_{ts}\left[\begin{array}{cc}
R & \bar{R}\\
R & \bar{R}
\end{array}\right]&\nonumber \\
   & (\lambda_{t}^{(-)}(\bar{R},\, R))^{-7}\, a_{ts^{\prime}}\left[\begin{array}{cc}
\bar{R} & R\\
R & \bar{R}
\end{array}\right]\,(\lambda_{s^{\prime}}^{(+)}(R,\, R))^{-1}.&
\end{flalign}
For framed knots $\mathcal{K}$ with framing number $p$, the invariants will be related to the zero-framed knot invariants as 
\begin{equation}
V_{R}^{\{U(N)\}}[\mathcal{K};\,p]=q^{p \kappa_R} V_{R}^{\{U(N)\}}[\mathcal{K};\,0]~.
\end{equation}
From the formal expression for these invariants, we see that they  involve  two types  of quantum $SU(N)$ Racah coefficients: (1) $ a_{ts}\left[\begin{array}{cc}
R & \bar{R}\\
R & \bar{R}
\end{array}\right]$  which is a symmetric matrix and (2) $a_{uv}\left[\begin{array}{cc}
R & R\\
\bar{R} & \bar{R}
\end{array}\right]$  which are known only for $R=\one,\twohor$ and \twover \cite{Zodinmawia:2011ud}.
It is a challenging problem to find the  Racah coefficients  for other representations. 

\subsection{Non Torus Links}
In the context of links, we can place different representations on the component knots. The invariants
are hence called multicolored links. We present the $U(N)$ Chern-Simons invariants for the 
two-component non-torus links(with framing zero)  in Figure \ref{fig:links}.
\begin{flalign}
V_{(R_{1},R_{2})}^{\{U(N)\}}[6_{2};\,0,0] = & q^{\frac{3l_{R_{1}}l_{R_{2}}}{N}}\sum_{s,t,s^{\prime}}\epsilon_{s}^{R_{1},R_{2}}\,\sqrt{dim_{q}s}\,\epsilon_{s^{\prime}}^{\bar{R}_{1},\bar{R}_{2}}\,\sqrt{dim_{q}s^{\prime}}\,(\lambda_{s}^{(+)}(R_{1},\, R_{2}))^{-3}\, a_{ts}\left[\begin{array}{cc}
R_{2} & \bar{R}_{1}\\
\bar{R}_{2} & R_{1}
\end{array}\right]&\nonumber \\
   & (\lambda_{t}^{(-)}(\bar{R}_{1},\, R_{2}))^{-2}\, a_{ts^{\prime}}\left[\begin{array}{cc}
\bar{R}_{1} & R_{2}\\
R_{1} & \bar{R}_{2}
\end{array}\right]\,(\lambda_{s^{\prime}}^{(+)}(\bar{R}_{1},\,\bar{R}_{2}))^{-1}.&
\end{flalign}
\begin{flalign}
V_{(R_{1},R_{2})}^{\{U(N)\}}[6_{3};\,0,0] = & q^{(-2\kappa_{R_{2}}+\frac{l_{R_{2}}^{2}}{N}-\frac{2l_{R_{1}}l_{R_{2}}}{N})}\sum_{s,t,s^{\prime},u,v}\epsilon_{s}^{R_{1},R_{2}}\,\sqrt{dim_{q}s}\,\epsilon_{v}^{\bar{R}_{1},\bar{R}_{2}}\,\sqrt{dim_{q}v}\,\lambda_{s}^{(+)}(R_{1},\, R_{2})&\nonumber \\
   & a_{ts}\left[\begin{array}{cc}
\bar{R}_{1} & R_{2}\\
R_{1} & \bar{R}_{2}
\end{array}\right]\,\lambda_{\bar{t}}^{(-)}(R_{1},\,\bar{R}_{2})\, a_{ts^{\prime}}\left[\begin{array}{cc}
\bar{R}_{1} & R_{2}\\
\bar{R}_{2} & R_{1}
\end{array}\right]\,(\lambda_{s^{\prime}}^{(-)}(R_{2},\,\bar{R}_{2}))^{-2}&\nonumber \\
   & a_{us^{\prime}}\left[\begin{array}{cc}
\bar{R}_{1} & R_{2}\\
\bar{R}_{2} & R_{1}
\end{array}\right]\,\lambda_{u}^{(-)}(\bar{R}_{1},\, R_{2})\, a_{uv}\left[\begin{array}{cc}
R_{2} & \bar{R}_{1}\\
\bar{R}_{2} & R_{1}
\end{array}\right]\,\lambda_{v}^{(+)}(\bar{R}_{1},\,\bar{R}_{2}).&
\end{flalign}
\begin{flalign}
V_{(R_{1},R_{2})}^{\{U(N)\}}[7_{1};\,0,0] = & q^{(-\kappa_{R_{2}}+\frac{l_{R_{2}}^{2}}{2N}+\frac{l_{R_{1}}l_{R_{2}}}{N})}\sum_{s,t,s^{\prime},u,v}\epsilon_{s}^{R_{1},R_{2}}\,\sqrt{dim_{q}s}\,\epsilon_{v}^{\bar{R}_{1},R_{2}}\,\sqrt{dim_{q}v}\,(\lambda_{s}^{(+)}(R_{1},\, R_{2}))&\nonumber \\
  & a_{ts}\left[\begin{array}{cc}
\bar{R}_{1} & R_{2}\\
R_{1} & \bar{R}_{2}
\end{array}\right]\,\lambda_{\bar{t}}^{(-)}(R_{1},\,\bar{R}_{2})\, a_{ts^{\prime}}\left[\begin{array}{cc}
\bar{R}_{1} & R_{2}\\
\bar{R}_{2} & R_{1}
\end{array}\right]\,(\lambda_{s^{\prime}}^{(-)}(R_{2},\,\bar{R}_{2}))^{-1}&\nonumber \\
   & a_{us^{\prime}}\left[\begin{array}{cc}
\bar{R}_{1} & \bar{R}_{2}\\
R_{2} & R_{1}
\end{array}\right]\,\lambda_{u}^{(+)}(\bar{R}_{1},\,\bar{R}_{2})^{-3}\, a_{uv}\left[\begin{array}{cc}
\bar{R}_{2} & \bar{R}_{1}\\
R_{2} & R_{1}
\end{array}\right]\,(\lambda_{v}^{(-)}(\bar{R}_{1},\, R_{2}))^{-1}.&
\end{flalign}
\begin{flalign}
V_{(R_{1},R_{2})}^{\{U(N)\}}[7_{2};\,0,0]= & q^{(-\kappa_{R_{2}}+\frac{l_{R_{2}}^{2}}{2N}-\frac{l_{R_{1}}l_{R_{2}}}{N})}\sum_{s,t,s^{\prime},u,v}\epsilon_{s}^{R_{1},R_{2}}\,\sqrt{dim_{q}s}\,\epsilon_{v}^{\bar{R}_{1},R_{2}}\,\sqrt{dim_{q}v}\,(\lambda_{s}^{(+)}(R_{1},\, R_{2}))^{-2}&\nonumber \\
   & a_{ts}\left[\begin{array}{cc}
\bar{R}_{1} & R_{1}\\
R_{2} & \bar{R}_{2}
\end{array}\right]\,(\lambda_{\bar{t}}^{(-)}(R_{2},\,\bar{R}_{2}))^{-1}\, a_{ts^{\prime}}\left[\begin{array}{cc}
\bar{R}_{1} & R_{1}\\
\bar{R}_{2} & R_{2}
\end{array}\right]\,(\lambda_{s^{\prime}}^{(-)}(R_{1},\,\bar{R}_{2}))&\nonumber \\
   & a_{us^{\prime}}\left[\begin{array}{cc}
\bar{R}_{1} & \bar{R}_{2}\\
R_{1} & R_{2}
\end{array}\right]\,\lambda_{\bar{u}}^{(+)}(R_{1},\, R_{2})^{2}\, a_{uv}\left[\begin{array}{cc}
\bar{R}_{2} & \bar{R}_{1}\\
R_{2} & R_{1}
\end{array}\right]\,\lambda_{v}^{(-)}(\bar{R}_{1},\, R_{2}).&
\end{flalign}
\begin{flalign}
V_{(R_{1},R_{2})}^{\{U(N)\}}[7_{3};\,0,0] = & q^{(-3\kappa_{R_{2}}+\frac{3l_{R_{2}}^{2}}{2N})}\sum_{s,t,s^{\prime},u,v}\epsilon_{s}^{R_{1},R_{2}}\,\sqrt{dim_{q}s}\,\epsilon_{v}^{\bar{R}_{1},R_{2}}\,\sqrt{dim_{q}v}\,\lambda_{s}^{(+)}(R_{1},\, R_{2})&\nonumber \\
  & a_{ts}\left[\begin{array}{cc}
\bar{R}_{1} & R_{2}\\
R_{1} & \bar{R}_{2}
\end{array}\right]\,\lambda_{t}^{(-)}(R_{1},\,\bar{R}_{2})\, a_{ts^{\prime}}\left[\begin{array}{cc}
\bar{R}_{1} & R_{2}\\
\bar{R}_{2} & R_{1}
\end{array}\right]\,(\lambda_{s^{\prime}}^{(-)}(R_{2},\,\bar{R}_{2}))^{-3}&\nonumber \\
   & a_{us^{\prime}}\left[\begin{array}{cc}
\bar{R}_{1} & \bar{R}_{2}\\
R_{2} & R_{1}
\end{array}\right]\,(\lambda_{u}^{(+)}(\bar{R}_{1},\,\bar{R}_{2}))^{-1}\, a_{uv}\left[\begin{array}{cc}
\bar{R}_{2} & \bar{R}_{1}\\
R_{2} & R_{1}
\end{array}\right]&\nonumber \\
   & (\lambda_{v}^{(-)}(\bar{R}_{1},\, R_{2}))^{-1}.&
\end{flalign}
Including the framing numbers $p_1,p_2$ on the component knots of these two-component torus links ${\cal L}$, the  framed multicolored invariant will be
\begin{equation}
V_{(R_{1},R_{2})}^{\{U(N)\}}[{\cal L} ;\,p_1,p_2]=q^{(p_1 \kappa_{R_1}+ p_2 \kappa_{R_2})}~ 
V_{(R_{1},R_{2})}^{\{U(N)\}}[{\cal L};\,0,0]~.
\end{equation}
Even though the formula denoting the  invariants of all these non-torus knots and links are available, we 
cannot write the explicit polynomial form for any representation placed on component knots because
Racah coefficients is known only for certain representations.  So, in the following section, we 
shall write the polynomial form of these invariants for those representations whose 
Racah coefficients are known \cite{Zodinmawia:2011ud}.
\section{Knot  Polynomials }
\label{sec:3}
\Yboxdim5pt
In this section we  present  the polynomial form  of  the $U(N)$ link invariant for non torus knots
in Figure \ref{fig:knots} for representation  whose Young tableau diagrams are $\yng(1)$ and $\yng(2)$. The polynomial corresponding to  representation $\yng(1)$ is proportional to HOMFLY-PT  polynomial $P(\lambda,t)[K]$ \cite{Freyd:1985dx,PT}
upto unknot $U$  normalisation:
\Yboxdim4pt
\begin{equation}
P(\lambda,q )[K] ={V_{\yng(1)}^{U(N)}[K;0]\over V_{\yng(1)}^{U(N)}[U]}= {(q^{1/2}-q^{-1/2})\over (\lambda^{1/2}-\lambda^{-1/2})}V_{\yng(1)}^{U(N)}[K;0]~.
\end{equation}
We list them so that we can directly use  them in the computation of 
reformulated invariants in section \ref{sec:5}.

\renewcommand{\baselinestretch}{1.5}\selectfont

\begin{enumerate}

\item
For fundamental representation \Yboxdim5pt$R=\yng(1)$ $~$placed on the knot, the $U(N)$ knot polynomials are
\Yboxdim4pt
\begin{center}
\begin{tabular}{c p{11cm}}
$V_{\yng(1)}^{U(N)}[4_{1}]=$ &$\frac{(\lambda-1)}{\lambda^{3/2}(q-1)\sqrt{q}}\big[-\lambda-\lambda q^{2}+\left(\lambda^{2}+\lambda+1\right)q\bigl]$
\end{tabular}  
\end{center}
\begin{center}
\begin{tabular}{c p{11cm}}
$V_{\yng(1)}^{U(N)}[5_{2}] =$ &$\frac{1}{(-1+q)\sqrt{q}\lambda^{7/2}}\bigl[q-q\lambda^{3}+\lambda(-1+\lambda^{2})+q^{2}\lambda(-1+\lambda^{2})\bigl]$
\end{tabular}  
\end{center}
\begin{center}
\begin{tabular}{c p{11cm}}
$V_{\yng(1)}^{U(N)}[6_{1}] =$ &$ \frac{1}{\lambda^{5/2}(q-1)\sqrt{q}}\bigl[-\lambda^{3}+\lambda+(\lambda-\lambda^{3})q^{2}+(\lambda^{4}+\lambda^{3}-\lambda^{2}-1)q\bigl]$
\end{tabular}  
\end{center}
\begin{center}
\begin{tabular}{c p{11cm}}
$V_{\yng(1)}^{U(N)}[6_{2}]  =$ & $\frac{(-1+\lambda)}{(-1+q)q^{3/2}\lambda^{5/2}}\bigl[-\lambda-q^{4}\lambda-q^{2}(1+2\lambda)+q(1+\lambda+\lambda^{2})+q^{3}(1+\lambda+\lambda^{2})\bigl]$
\end{tabular}  
\end{center}
\begin{center}
\begin{tabular}{c p{11cm}}
$V_{\yng(1)}^{U(N)}[6_{3}]=$ & $-\frac{(-1+\lambda)}{(-1+q)q^{3/2}\lambda^{3/2}}\bigl[-\lambda-q^{4}\lambda+q(1+\lambda+\lambda^{2})+q^{3}(1+\lambda+\lambda^{2})-q^{2}(1+3\lambda+\lambda^{2})\bigl]$
\end{tabular}  
\end{center}
 \begin{center}
\begin{tabular}{c p{11cm}} 
$V_{\yng(1)}^{U(N)}[7_{2}]=$ & $\frac{1}{(-1+q)\sqrt{q}\lambda^{9/2}}\bigl[\lambda(-1+\lambda^{3})+q^{2}\lambda(-1+\lambda^{3})-q(-1-\lambda^{2}+\lambda^{3}+\lambda^{4})\bigl]$
\end{tabular}  
\end{center}
\begin{center}
\begin{tabular}{c p{11cm}}
$V_{\yng(1)}^{U(N)}[7_{3}] =$ & $\frac{\lambda^{3/2}}{(-1+q)q^{3/2}}\bigl[-1+q+\lambda^{2}-q\lambda^{3}+q^{4}(-1+\lambda^{2})+q^{2}(-1-\lambda+2\lambda^{2})-q^{3}(-1+\lambda^{3})\bigl]$
\end{tabular}  
\end{center}
\begin{center}
\begin{tabular}{c p{11cm}}
$V_{\yng(1)}^{U(N)}[7_{4}]=$ & $\frac{(-1+\lambda)\lambda^{1/2}}{(-1+q)q^{1/2}}\bigl[(1+\lambda)^{2}+q^{2}(1+\lambda)^{2}-q(2+2\lambda+2\lambda^{2}+\lambda^{3})\bigl]$
\end{tabular}  
\end{center}
\begin{center}
\begin{tabular}{c p{11cm}}
$V_{\yng(1)}^{U(N)}[7_{5}]  =$ & $\ensuremath{\frac{(-1+\lambda)}{(-1+q)q^{3/2}\lambda^{9/2}}\bigl[\lambda(1+\lambda)+q^{4}\lambda(1+\lambda)-q(1+\lambda)^{2}}-q^{3}(1+\lambda)^{2}+q^{2}(1+2\lambda+2\lambda^{2})\bigl]$
\end{tabular}  
\end{center}
\begin{center}
\begin{tabular}{c p{11cm}}
$V_{\yng(1)}^{U(N)}[7_{6}]  =$ & $-\frac{(-1+\lambda)}{(-1+q)q^{3/2}\lambda^{7/2}}\bigl[\lambda^{2}+q^{4}\lambda^{2}-q\lambda(2+2\lambda+\lambda^{2})-q^{3}\lambda(2+2\lambda+\lambda^{2})+q^{2}(1+2\lambda+3\lambda^{2}+\lambda^{3})\bigl]$
\end{tabular}  
\end{center}
\begin{center}
\begin{tabular}{c p{11cm}}
$V_{\yng(1)}^{U(N)}[7_{7}]=$ & $\frac{(-1+\lambda)}{(-1+q)q^{3/2}\lambda^{3/2}}\bigl[\lambda+q^{4}\lambda-q(1+2\lambda+2\lambda^{2})-q^{3}(1+2\lambda+2\lambda^{2})+q^{2}(2+4\lambda+2\lambda^{2}+\lambda^{3})\bigl]$
\end{tabular}  
\end{center}
\begin{center}
\begin{tabular}{c p{11cm}} 
$V_{\yng(1)}^{U(N)}[8_{1}]=$ &$ \frac{1}{(q-1)\sqrt{q}\lambda^{7/2}}\bigl[-\lambda^{4}+\lambda+q^{2}(\lambda-\lambda^{4})+q(\lambda^{5}+\lambda^{4}-\lambda^{2}-1)\bigl]$
\end{tabular}  
\end{center}
\begin{center}
\begin{tabular}{c p{11cm}}
$V_{\yng(1)}^{U(N)}[9_{1}]=$
&$\frac{1}{(q-1)\sqrt{q}\lambda^{11/2}}\bigl[\lambda(\lambda^{4}-1)q^{2}-(\lambda^{5}+\lambda^{4}-\lambda^{2}-1)q+\lambda(\lambda^{4}-1)\bigl]$
\end{tabular}  
\end{center}
\begin{center}
\begin{tabular}{c p{11cm}}
$V_{\yng(1)}^{U(N)}[10_{1}]=$
& $\frac{1}{(q-1)\sqrt{q}\lambda^{9/2}}\bigl[-\lambda^{5}+\lambda+q^{2}(\lambda-\lambda^{5})+q(\lambda^{6}+\lambda^{5}-\lambda^{2}-1)\bigl]$
\end{tabular}  
\end{center}
\Yboxdim5pt
\item For symmetric second rank representation $R=\yng(2)$ , the knot polynomials are
\Yboxdim4pt
\begin{center}
\begin{tabular}{c p{11cm}}
$V_{\yng(2)}^{U(N)}[4_{1}]=$ &$\frac{(-1+\lambda ) (-1+q \lambda ) }{(-1+q)^2 q^2 (1+q) \lambda ^3}\bigl[(-1+\lambda ) \lambda +3 q^3 \lambda ^2-q^6 (-1+\lambda ) \lambda ^2+q^4 \lambda  (-1+\lambda ^2)+q^5 \lambda ^2 (-1-\lambda +\lambda ^2)-q (-1+\lambda +\lambda ^2)+q^2 (\lambda -\lambda ^3)\bigl]$
\end{tabular}  
\end{center}
\begin{center}
\begin{tabular}{c p{11cm}}
$V_{\yng(2)}^{U(N)}[5_{2}]=$ &$\frac{1}{(-1+q)^2 q^5 (1+q) \lambda ^7}\bigl[q (-1+\lambda )^2-(-1+\lambda )^2 \lambda +q^9 \lambda ^4 (-1+\lambda ^2)+q^3 (-1+\lambda )^2 \lambda ^2 (1+\lambda +\lambda ^2)+q^6 \lambda ^4 (-1-\lambda +2 \lambda ^2)+q^5 \lambda ^2 (-1+\lambda -\lambda ^2+\lambda ^3)-q^8 \lambda ^3 (-1+\lambda -\lambda ^2+\lambda ^3)+q^4 \lambda  (-1+\lambda +\lambda ^2-\lambda ^5)+q^7 (\lambda ^3-\lambda ^6)\bigl]$
\end{tabular}  
\end{center}
\begin{center}
\begin{tabular}{c p{11cm}}
$V_{\yng(2)}^{U(N)}[6_{1}]=$ &$\frac{(-1+\lambda ) (-1+q \lambda ) }{(-1+q)^2 q^4 (1+q) \lambda ^5}\bigl[(-1+\lambda ) \lambda +q (-1+\lambda )^2 (1+\lambda )+q^2 (-1+\lambda )^2 \lambda  (1+\lambda )+q^7 \lambda ^4 (-3+\lambda ^2)+2 q^6 \lambda ^3 (-1+\lambda ^2)-2 q^3 \lambda ^2 (-1+\lambda +\lambda ^2)+q^5 \lambda ^2 (-1+2 \lambda +4 \lambda ^2)-q^4 \lambda  (1-\lambda -3 \lambda ^2+2 \lambda ^3+\lambda ^4)+q^8 (\lambda ^3-\lambda ^5)\bigl]$
\end{tabular}  
\end{center}
\begin{center}
\begin{tabular}{c p{11cm}}
$V_{\yng(2)}^{U(N)}[6_{2}]=$ &$\frac{(-1+\lambda ) (-1+q \lambda ) }{(-1+q)^2 q^6 (1+q) \lambda ^5}\bigl[q+(-1+\lambda ) \lambda -q \lambda ^2-q^9 (-4+\lambda ) \lambda ^2-q^{12} (-1+\lambda ) \lambda ^2-q^2 (-1+\lambda )^2 (1+\lambda )+q^3 \lambda  (-3+2 \lambda )-2 q^4 (-1+\lambda ^2)+q^{10} \lambda  (-1+\lambda ^2)+q^{11} \lambda ^2 (-1-\lambda +\lambda ^2)+q^6 (-1-3 \lambda +4 \lambda ^2)+q^8 \lambda  (2-3 \lambda ^2+\lambda ^3)+q^7 (1-2 \lambda -3 \lambda ^2+\lambda ^4)+q^5 (-1+3 \lambda +2 \lambda ^2-2 \lambda ^3+\lambda ^4)\bigl]$
\end{tabular}  
\end{center}
\begin{center}
\begin{tabular}{c p{11cm}}
$V_{\yng(2)}^{U(N)}[6_{3}]=$ &$-\frac{(-1+\lambda ) (-1+q \lambda ) }{(-1+q)^2 q^5 (1+q) \lambda ^3}\bigl[-(-1+\lambda ) \lambda +q^{12} (-1+\lambda ) \lambda ^2+q^3 (1+3 \lambda -4 \lambda ^2)+q (-1+\lambda ^2)+q^9 \lambda ^2 (-4+3 \lambda +\lambda ^2)+q^7 (-1+4 \lambda +\lambda ^2-4 \lambda ^3)+q^5 \lambda  (-4+\lambda +4 \lambda ^2-\lambda ^3)-q^4 (2-3 \lambda -3 \lambda ^2+\lambda ^3)+q^{10} \lambda  (1-\lambda -2 \lambda ^2+\lambda ^3)+q^2 (1-2 \lambda -\lambda ^2+\lambda ^3)-q^8 \lambda (1-3 \lambda -3 \lambda ^2+2 \lambda ^3)+q^{11} (\lambda ^2-\lambda ^4)+q^6 (1+\lambda -9 \lambda ^2+\lambda ^3+\lambda ^4)\bigl]$
\end{tabular}  
\end{center}
\begin{center}
\begin{tabular}{c p{11cm}}
$V_{\yng(2)}^{U(N)}[7_{2}]=$ &$-\frac{(-1+\lambda )^2 }{(-1+q) q^7 \lambda ^9}\bigl[q-\lambda -q^2 \lambda -2 q^4 \lambda +q^3 (1+\lambda ^2)-q^8 \lambda  (1+\lambda +2 \lambda ^2+2 \lambda ^3+\lambda ^4)-q^6 \lambda  (2+\lambda +3 \lambda ^2+3 \lambda ^3+3 \lambda ^4+\lambda ^5)+q^5 (1+\lambda ^2+\lambda ^4+\lambda ^5+\lambda ^6)+q^7 (1+\lambda +3 \lambda ^2+3 \lambda ^3+4 \lambda ^4+3 \lambda ^5+\lambda ^6)\bigl]$
\end{tabular}  
\end{center}
\begin{center}
\begin{tabular}{c p{11cm}}
$V_{\yng(2)}^{U(N)}[7_{3}]=$ &$\frac{(-1+\lambda ) \lambda ^3 (-1+q \lambda ) }{(-1+q)^2 q^3 (1+q)}\bigl[1+q (-1+\lambda )-q^{12} (-1+\lambda )+\lambda -q^{14} (-1+\lambda ) \lambda ^2+q^{13} (-1+\lambda )^2 \lambda  (1+\lambda )-q^2 (1+2 \lambda )-q^4 \lambda  (-4+\lambda ^2)-2 q^3 (-1+\lambda ^2)+q^5 (-2-\lambda +3 \lambda ^2)+q^8 (-1+\lambda +4 \lambda ^2-2 \lambda ^3)+q^9 (1-2 \lambda +\lambda ^3)+q^6 (1-3 \lambda -2 \lambda ^2+\lambda ^3)+q^{10} \lambda  (2-2 \lambda -\lambda ^2+\lambda ^3)+q^7 (1+3 \lambda -3 \lambda ^2-2 \lambda ^3+\lambda ^4)+q^{11} (-1+2 \lambda ^2-2 \lambda ^3+\lambda ^4)\bigl]$
\end{tabular}  
\end{center}
\begin{center}
\begin{tabular}{c p{11cm}}
$V_{\yng(2)}^{U(N)}[7_{4}]=$ &$\frac{(-1+\lambda ) \lambda  (-1+q \lambda )}{(-1+q)^2 q (1+q)}\bigl[-q^{10} (-1+\lambda ) \lambda ^4+(1+\lambda )^2+2 q (-1+\lambda ^2)+q^2 (-1-6 \lambda -\lambda ^2+\lambda ^3)+q^9 \lambda ^3 (2-2 \lambda -\lambda ^2+\lambda ^3)-q^3 (-4-2 \lambda +5 \lambda ^2+\lambda ^3)+q^8 \lambda ^2 (3-2 \lambda -3 \lambda ^2+2 \lambda ^3)+q^5 (-2-4 \lambda +6 \lambda ^2+3 \lambda ^3-3 \lambda ^4)+q^7 \lambda  (2-2 \lambda -4 \lambda ^2+3 \lambda ^3+\lambda ^4)-q^4 (1-6 \lambda -4 \lambda ^2+4 \lambda ^3+\lambda ^4)+q^6 (1-2 \lambda -4 \lambda ^2+5 \lambda ^3+\lambda ^4-\lambda ^5)\bigl]$
\end{tabular}  
\end{center}
\begin{center}
\begin{tabular}{c p{11cm}}
$V_{\yng(2)}^{U(N)}[7_{5}]=$ &$==\frac{(-1+\lambda ) (-1+q \lambda )}{(-1+q)^2 q^9 (1+q) \lambda ^9}\bigl[(-1+\lambda ) \lambda -q^{13} (-1+\lambda ) \lambda ^3+q^2 (-1+\lambda )^3 (1+\lambda )+q^{14} \lambda ^3 (1+\lambda )-q^{12} \lambda ^3 (3+\lambda )+3 q^{11} \lambda ^2 (-1+\lambda ^2)+q^{10} \lambda  (-1+6 \lambda ^2)-q^3 \lambda  (3-4 \lambda +\lambda ^3)+q (1-2 \lambda ^2+\lambda ^3)+q^8 \lambda  (3-5 \lambda -5 \lambda ^2+3 \lambda ^3)+q^9 (\lambda +5 \lambda ^2-3 \lambda ^3-3 \lambda ^4)+q^4 (2-\lambda -5 \lambda ^2+3 \lambda ^3+\lambda ^4)+q^7 (1-3 \lambda -5 \lambda ^2+6 \lambda ^3+\lambda ^4)-q^6 (1+2 \lambda -7 \lambda ^2+2 \lambda ^4)+q^5 (-1+5 \lambda -\lambda ^2-5 \lambda ^3+2 \lambda ^4)\bigl]$
\end{tabular}  
\end{center}
\begin{center}
\begin{tabular}{c p{11cm}}
$V_{\yng(2)}^{U(N)}[7_{6}]=$ &$\frac{(-1+\lambda ) (-1+q \lambda )}{(-1+q)^2 q^6 (1+q) \lambda ^7}\bigl[(-1+\lambda )^2 \lambda ^2-q^{12} (-1+\lambda ) \lambda ^4+q^{11} \lambda ^4 (-2+\lambda ^2)+q \lambda  (-2+3 \lambda +\lambda ^2-2 \lambda ^3)-q^{10} \lambda ^3 (2-3 \lambda ^2+\lambda ^3)-q^9 \lambda ^3 (-1-7 \lambda +2 \lambda ^2+\lambda ^3)+q^7 \lambda ^2 (3-7 \lambda -7 \lambda ^2+4 \lambda ^3)+q^6 \lambda ^2 (-4-6 \lambda +10 \lambda ^2+\lambda ^3-\lambda ^4)+q^4 \lambda  (-1+8 \lambda -\lambda ^2-7 \lambda ^3+\lambda ^4)+q^3 \lambda  (2-8 \lambda ^2+4 \lambda ^3+\lambda ^4)+q^8 \lambda ^2 (1+5 \lambda -4 \lambda ^2-4 \lambda ^3+2 \lambda ^4)+q^5 \lambda  (-2-2 \lambda +10 \lambda ^2+2 \lambda ^3-4 \lambda ^4+\lambda ^5)-q^2 (-1+\lambda +4 \lambda ^2-3 \lambda ^3-2 \lambda ^4+\lambda ^5)\bigl]$
\end{tabular}  
\end{center}
\begin{center}
\begin{tabular}{c p{11cm}}
$V_{\yng(2)}^{U(N)}[7_{7}]=$ &$\frac{(-1+q \lambda )}{(-1+q)^2 q^5 (1+q) \lambda ^3} \bigl[(-1+\lambda )^2 \lambda +q^{12} (-1+\lambda )^3 \lambda ^2-q (-1+\lambda )^2 (1+2 \lambda )-2 q^{11} (-1+\lambda )^2 \lambda ^2 (-1-\lambda +\lambda ^2)+q^3 (-1+\lambda )^2 (1+7 \lambda +2 \lambda ^2)+q^5 (-1+\lambda )^2 (1-9 \lambda -8 \lambda ^2+2 \lambda ^3)+q^9 (-1+\lambda )^2 \lambda  (-1-8 \lambda -\lambda ^2+2 \lambda ^3)+q^2 (2-5 \lambda +\lambda ^2+4 \lambda ^3-2 \lambda ^4)-q^7 (-1+\lambda )^2 (1-5 \lambda -11 \lambda ^2+2 \lambda ^4)+q^8 \lambda  (-3+11 \lambda -16 \lambda ^3+8 \lambda ^4)+q^6 (2-18 \lambda ^2+15 \lambda ^3+5 \lambda ^4-4 \lambda ^5)+q^4 (-4+6 \lambda +8 \lambda ^2-12 \lambda ^3+\lambda ^4+\lambda ^5)+q^{10} \lambda  (1-2 \lambda -4 \lambda ^2+8 \lambda ^3-\lambda ^4-3 \lambda ^5+\lambda ^6)\bigl]$
\end{tabular}  
\end{center}
\begin{center}
\begin{tabular}{c p{11cm}}
$V_{\yng(2)}^{U(N)}[8_{1}]=$ &$\frac{1}{(-1+q)^2 q^6 (1+q) \lambda ^7}\bigl[q (-1+\lambda )^2-(-1+\lambda )^2 \lambda -q^4 (-1+\lambda )^2 \lambda +q^3 (-1+\lambda )^2 \lambda ^2+q^5 \lambda ^5 (-1+\lambda ^3)+q^{11} \lambda ^5 (-1+\lambda +\lambda ^3-\lambda ^4)-q^6 \lambda ^4 (-1+\lambda -\lambda ^2+\lambda ^4)+q^9 \lambda ^5 (-1+\lambda -\lambda ^2+\lambda ^4)+q^8 \lambda ^4 (-1-\lambda +2 \lambda ^4)+q^{10} \lambda ^4 (1-\lambda +2 \lambda ^2-\lambda ^3-\lambda ^4-\lambda ^5+\lambda ^6)+q^7 (\lambda ^4+\lambda ^6-\lambda ^8-\lambda ^9)\bigl]$
\end{tabular}  
\end{center}
\begin{center}
\begin{tabular}{c p{11cm}}
$V_{\yng(2)}^{U(N)}[9_{2}]=$ & $\frac{1}{(-1+q)^2 q^9 (1+q) \lambda ^{11}}\bigl[q (-1+\lambda )^2-(-1+\lambda )^2 \lambda -q^4 (-1+\lambda )^2 \lambda +q^3 (-1+\lambda )^2 \lambda ^2+q^{13} \lambda ^6 (-1+\lambda ^4)+q^{10} \lambda ^6 (-1-\lambda +2 \lambda ^4)+q^9 \lambda ^4 (-1+\lambda -\lambda ^2+\lambda ^5)-q^{12} \lambda ^5 (-1+\lambda -\lambda ^2+\lambda ^5)+q^7 \lambda ^5 (1-\lambda -\lambda ^4+\lambda ^5)-q^8 \lambda ^4 (1-2 \lambda +\lambda ^2-\lambda ^3-\lambda ^4+\lambda ^5+\lambda ^6)+q^{11} (\lambda ^5+\lambda ^7-\lambda ^9-\lambda ^{10})\bigl]$
\end{tabular}  
\end{center}
\begin{center}
\begin{tabular}{c p{11cm}}
$V_{\yng(2)}^{U(N)}[10_{2}]=$ &$\frac{1}{(-1+q)^2 q^8 (1+q) \lambda ^9}\bigl[q (-1+\lambda )^2-(-1+\lambda )^2 \lambda -q^4 (-1+\lambda )^2 \lambda +q^3 (-1+\lambda )^2 \lambda ^2+q^7 \lambda ^6 (-1+\lambda ^4)+q^{13} \lambda ^6 (-1+\lambda +\lambda ^4-\lambda ^5)-q^8 \lambda ^5 (-1+\lambda -\lambda ^2+\lambda ^5)+q^{11} \lambda ^6 (-1+\lambda -\lambda ^2+\lambda ^5)+q^{10} \lambda ^5 (-1-\lambda +2 \lambda ^5)-q^9 \lambda ^5 (-1-\lambda ^2+\lambda ^5+\lambda ^6)+q^{12} \lambda ^5 (1-\lambda +2 \lambda ^2-\lambda ^3-\lambda ^5-\lambda ^6+\lambda ^7)\bigl]$
\end{tabular}  
\end{center}
\end{enumerate}
\renewcommand{\baselinestretch}{1}\selectfont
There seems to be a symmetry transformation on the polynomial variables which gives the 
$U(N)$ invariants of knots carrying antisymmetric second rank tensor representation $R=\yng(1,1)$.
The symmetry relation for these non-torus knots (see also equation (7) in \cite{Itoyama:2012fq})
\begin{align}
\Yboxdim4pt
V_{\yng(2)}^{U(N)}[\mathcal{K}](q^{-1},\lambda)=V_{\yng(1,1)}^{U(N)}[\mathcal{K}](q,\lambda). \label{symm}
\end{align}
Before we use these polynomial invariants in verifying Ooguri-Vafa conjecture, we shall enumerate
the multi-colored link polynomials in the following section.
\section{Link Polynomials}
\label{sec:4}
\Yboxdim5pt
In this section we list the $U(N)$ link invariant for non torus links
given in Figure \ref{fig:links} for representations $R_1,R_2\in   \{\yng(1), \yng(2), \yng(1,1)\}$.

\renewcommand{\baselinestretch}{1.5}\selectfont
\begin{enumerate}
\item
For $R_1=\yng(1)$, $R_2=\yng(1)$    : 
\Yboxdim4pt
\begin{center}
\begin{tabular}{c p{11cm}}
$V_{(\yng(1),\,\yng(1))}^{U(N)}[6_{2}]=$ &$\frac{1}{(-1+q)^2 q \lambda }\bigl[\lambda  \left(-1+\lambda ^2\right)+q^4 \lambda  \left(-1+\lambda ^2\right)+q \left(1+\lambda -2 \lambda ^3\right)+q^3 \left(1+\lambda -2 \lambda ^3\right)+q^2 \left(-1-2 \lambda +\lambda ^2+2 \lambda ^3\right)\bigl]$
\end{tabular}  
\end{center}
\begin{center}
\begin{tabular}{c p{11cm}}
$V_{(\yng(1),\,\yng(1))}^{U(N)}[6_{3}]=$ &$\frac{(-1+\lambda ) }{(-1+q)^2 q \lambda ^3}\bigl[\lambda +q^4 \lambda -q \left(1+3 \lambda +2 \lambda ^2\right)-q^3 \left(1+3 \lambda +2 \lambda ^2\right)+q^2 \left(2+4 \lambda +3 \lambda ^2+\lambda ^3\right)\bigl]$
\end{tabular}  
\end{center}
\begin{center}
\begin{tabular}{c p{11cm}}
$V_{(\yng(1),\,\yng(1))}^{U(N)}[7_{1}]=$ &$\frac{(-1+\lambda ) }{(-1+q)^2 q^2 \lambda ^2}\bigl[-\lambda -q^6 \lambda +q (1+\lambda )^2+q^5 (1+\lambda )^2-q^2 \left(2+3 \lambda +\lambda ^2\right)+q^3 \left(2+3 \lambda +\lambda ^2\right)-q^4 \left(2+3 \lambda +\lambda ^2\right)\bigl]$
\end{tabular}  
\end{center}
\begin{center}
\begin{tabular}{c p{11cm}}
$V_{(\yng(1),\,\yng(1))}^{U(N)}[7_{2}]=$ &$\frac{(-1+\lambda ) }{(-1+q)^2 q^2 \lambda ^2}\bigl(-\lambda -q^6 \lambda +q (1+\lambda )^2-2 q^2 (1+\lambda )^2-2 q^4 (1+\lambda )^2+q^5 (1+\lambda )^2+q^3 \left(2+5 \lambda +3 \lambda ^2\right)\bigl]$
\end{tabular}  
\end{center}
\begin{center}
\begin{tabular}{c p{11cm}}
$V_{(\yng(1),\,\yng(1))}^{U(N)}[7_{3}]=$ &$\frac{(-1+\lambda ) }{(-1+q)^2 q \lambda ^3}\bigl[-\lambda  (1+\lambda )-q^4 \lambda  (1+\lambda )+q (1+\lambda )^3+q^3 (1+\lambda )^3-q^2 \left(2+4 \lambda +5 \lambda ^2+\lambda ^3\right)\bigl]$
\end{tabular}  
\end{center}
\Yboxdim5pt
\item
For $R_1=\yng(1)$, $R_2=\yng(2)$    : 
\Yboxdim4pt
\begin{center}
\begin{tabular}{c p{11cm}}
$V_{(\yng(1),\,\yng(2))}^{U(N)}[6_{2}]=$ &$\frac{(-1+\lambda ) }{(-1+q)^3 \sqrt{q} (1+q) \lambda ^{3/2}}\bigl[q-\lambda -q^7 \lambda ^3+q^8 \lambda ^3+q^2 \lambda  (1+\lambda +\lambda ^2)-q^6 \lambda  (1+\lambda +\lambda ^2)+q^5 \lambda  (1+2 \lambda +2 \lambda ^2)-q^4 (-1+\lambda ^3)-q^3 (1+2 \lambda +\lambda ^2+\lambda ^3)\bigl]$
\end{tabular}  
\end{center}
\begin{center}
\begin{tabular}{c p{11cm}}
$V_{(\yng(1),\,\yng(2))}^{U(N)}[6_{3}]=$ &$\frac{(-1+\lambda ) }{(-1+q)^3 q^{5/2} (1+q) \lambda ^{7/2}}\bigl[-\lambda +q^7 \lambda ^2-q^2 (1+\lambda )+q^5 \lambda ^3 (1+\lambda )+q (1+\lambda )^2-q^6 \lambda  (1+\lambda +2 \lambda ^2)+q^4 (1+3 \lambda +2 \lambda ^2+\lambda ^3)-q^3 (1+2 \lambda +2 \lambda ^2+2 \lambda ^3)\bigl]$
\end{tabular}  
\end{center}
\begin{center}
\begin{tabular}{c p{11cm}}
$V_{(\yng(1),\,\yng(2))}^{U(N)}[7_{1}]=$ &$\frac{1}{(-1+q)^3 q^{5/2} (1+q) \lambda ^{5/2}}\bigl[q+q^2 (-1+\lambda )+(-1+\lambda ) \lambda -q^{10} (-1+\lambda ) \lambda ^2-q \lambda ^3-q^8 (-1+\lambda ) \lambda ^3+q^7 (1+\lambda -2 \lambda ^3)-2 q^4 (-1+\lambda ^3)+q^9 \lambda  (-1+\lambda ^3)+q^5 (-1+\lambda +\lambda ^3-\lambda ^4)+q^3 (-1-\lambda +\lambda ^2+\lambda ^4)+q^6 (-1-\lambda +\lambda ^3+\lambda ^4)\bigl]$
\end{tabular}  
\end{center}
\begin{center}
\begin{tabular}{c p{11cm}}
$V_{(\yng(1),\,\yng(2))}^{U(N)}[7_{2}]=$ &$-\frac{(-1+q \lambda ) }{(-1+q)^3 q^{7/2} (1+q) \lambda ^{5/2}}\bigl[(q+(-1+\lambda ) \lambda +q^9 (-1+\lambda ) \lambda -q \lambda ^2+q^5 (-1+3 \lambda +\lambda ^2-3 \lambda ^3)+q^8 (\lambda -\lambda ^3)-q^2 (1-2 \lambda +\lambda ^3)+q^4 (2+\lambda -4 \lambda ^2+\lambda ^3)+q^7 (1-2 \lambda ^2+\lambda ^3)+q^3 (-1-3 \lambda +3 \lambda ^2+\lambda ^3)+q^6 (-1-3 \lambda +3 \lambda ^2+\lambda ^3)\bigl]$
\end{tabular}  
\end{center}
\begin{center}
\begin{tabular}{c p{11cm}}
$V_{(\yng(1),\,\yng(2))}^{U(N)}[7_{3}]=$ &$\frac{1}{(-1+q)^3 q^{3/2} (1+q) \lambda ^{7/2}}\bigl[q^2 (-1+\lambda ^2)+\lambda  (-1+\lambda ^2)+q^4 (1+2 \lambda -3 \lambda ^4)+q^7 (\lambda ^2-\lambda ^4)-q (-1-\lambda +\lambda ^3+\lambda ^4)+q^6 \lambda  (-1-\lambda +\lambda ^3+\lambda ^4)+q^5 (\lambda ^3-\lambda ^5)+q^3 (-1-\lambda -2 \lambda ^2+2 \lambda ^3+\lambda ^4+\lambda ^5)\bigl]$
\end{tabular}  
\end{center}
\renewcommand{\baselinestretch}{1}\selectfont
Interchanging $R_1,R_2$ on the two components of the link,  gives the same polynomial:
\begin{equation}
V_{(\yng(1),\yng(2))}^{U(N)}[\mathcal{L}]=V_{(\yng(2),\yng(1))}^{U(N)}[\mathcal{L}]~,
\end{equation}
and replacing the second rank symmetric representation $\yng(2)$ by antisymmetric 
representation $\yng(1,1)$, the link polynomials are related as follows:

\begin{align}
\Yvcentermath1
V_{(\yng(1),\yng(2))}^{U(N)}[\mathcal{L}](q^{-1},\lambda)=-V_{\left(\yng(1),\yng(1,1)\right)}^{U(N)}[\mathcal{L}](q,\lambda)
\end{align}

\renewcommand{\baselinestretch}{1.5}\selectfont
\item
\Yboxdim5pt
For $R_1=\yng(2)$, $R_2=\yng(2)$   : 
\Yboxdim4pt
\begin{center}
\begin{tabular}{c p{11cm}}
$V_{(\yng(2),\,\yng(2))}^{U(N)}[6_{2}]=$ &$\frac{1}{(-1+q)^4 q (1+q)^2 \lambda ^2}\big[-q^2 (-1+\lambda )^2-(-1+\lambda )^2 \lambda +q (-1+\lambda )^2 (1+\lambda )+q^{15} \lambda ^4 (-1+\lambda ^2)-q^{13} \lambda ^4 (-2+\lambda +\lambda ^2)+q^3 (-1+\lambda )^2 \lambda  (-2+\lambda +\lambda ^2+\lambda ^3)+q^{12} \lambda ^3 (-2-\lambda -2 \lambda ^2+5 \lambda ^3)+q^{11} \lambda ^2 (-1+\lambda -2 \lambda ^2+4 \lambda ^3-2 \lambda ^4)+q^{10} \lambda  (-1+2 \lambda +\lambda ^2+3 \lambda ^3-5 \lambda ^5)+q^8 \lambda  (1-4 \lambda +\lambda ^2-3 \lambda ^3+4 \lambda ^4+\lambda ^5)+q^5 (-1+4 \lambda -4 \lambda ^2-\lambda ^4+2 \lambda ^5)+q^9 \lambda  (1+2 \lambda -3 \lambda ^2-5 \lambda ^4+5 \lambda ^5)+q^7 (1-4 \lambda +\lambda ^2+2 \lambda ^3+3 \lambda ^4+\lambda ^5-4 \lambda ^6)+q^{14} (\lambda ^3+\lambda ^5-2 \lambda ^6)+2 q^4 (1-2 \lambda +\lambda ^3+\lambda ^5-\lambda ^6)+q^6 (-1+\lambda +5 \lambda ^2-4 \lambda ^3+\lambda ^4-5 \lambda ^5+3 \lambda ^6)\bigl]$
\end{tabular}  
\end{center}
\begin{center}
\begin{tabular}{c p{11cm}}
$V_{(\yng(2),\,\yng(2))}^{U(N)}[6_{3}]=$ &$\frac{(-1+\lambda ) (-1+q \lambda )}{(-1+q)^4 q^7 (1+q)^2 \lambda ^6}\bigl[(-1+\lambda ) \lambda +q^{12} (-1+\lambda )^2 \lambda ^2+q (1+\lambda -3 \lambda ^2)+q^2 (-2+4 \lambda +\lambda ^2-2 \lambda ^3)+q^{11} \lambda ^2 (-3+2 \lambda +3 \lambda ^2-2 \lambda ^3)+q^3 (-1-5 \lambda +9 \lambda ^2+3 \lambda ^3)+q^8 \lambda  (2-13 \lambda -5 \lambda ^2+10 \lambda ^3)+q^6 (-2+\lambda +19 \lambda ^2-2 \lambda ^3-6 \lambda ^4)+q^4 (4-5 \lambda -11 \lambda ^2+5 \lambda ^3+\lambda ^4)+q^5 (-1+9 \lambda -8 \lambda ^2-12 \lambda ^3+2 \lambda ^4)+q^9 \lambda  (2+7 \lambda -10 \lambda ^2-3 \lambda ^3+3 \lambda ^4)+q^{10} \lambda  (-1+2 \lambda +6 \lambda ^2-6 \lambda ^3-\lambda ^4+\lambda ^5)-q^7 (-1+7 \lambda +2 \lambda ^2-17 \lambda ^3+\lambda ^4+2 \lambda ^5)\bigl]$
\end{tabular}  
\end{center}
\begin{center}
\begin{tabular}{c p{11cm}}
$V_{(\yng(2),\,\yng(2))}^{U(N)}[7_{1}]=$ &$\frac{(-1+\lambda ) (-1+q \lambda ) }{(-1+q)^4 q^6 (1+q)^2 \lambda ^4}\bigl[(-1+\lambda ) \lambda -q^{18} (-1+\lambda ) \lambda ^2+q (1+\lambda -2 \lambda ^2)+q^{17} \lambda ^2 (-2+\lambda ^2)+q^3 \lambda  (-4+3 \lambda +\lambda ^2)+q^2 (-2+2 \lambda +\lambda ^2-\lambda ^3)+q^7 (5+5 \lambda -7 \lambda ^2+\lambda ^3)+q^4 (3+2 \lambda -5 \lambda ^2+\lambda ^3)-q^{15} \lambda  (-1-6 \lambda +2 \lambda ^2+\lambda ^3)+q^{14} \lambda  (3-5 \lambda -4 \lambda ^2+2 \lambda ^3)+q^{13} (1-5 \lambda -7 \lambda ^2+5 \lambda ^3)+q^{12} (-2-3 \lambda +11 \lambda ^2+2 \lambda ^3-2 \lambda ^4)+q^6 (1-8 \lambda +3 \lambda ^2+\lambda ^3-\lambda ^4)+q^{11} (-1+10 \lambda +2 \lambda ^2-6 \lambda ^3+\lambda ^4)-q^{16} (\lambda -3 \lambda ^3+\lambda ^4)-q^9 (2+11 \lambda -5 \lambda ^2-3 \lambda ^3+\lambda ^4)+q^5 (-4+3 \lambda +2 \lambda ^2-3 \lambda ^3+\lambda ^4)+q^8 (-5+7 \lambda +6 \lambda ^2-3 \lambda ^3+\lambda ^4)+q^{10} (5-\lambda -12 \lambda ^2+\lambda ^3+\lambda ^4)\bigl]$
\end{tabular}  
\end{center}
\begin{center}
\begin{tabular}{c p{11cm}}
$V_{(\yng(2),\,\yng(2))}^{U(N)}[7_{2}]=$ &$\frac{(-1+\lambda ) (-1+q \lambda ) }{(-1+q)^4 q^8 (1+q)^2 \lambda ^4}\bigl[((-1+\lambda ) \lambda -q^{18} (-1+\lambda ) \lambda ^2+q (1+\lambda -2 \lambda ^2)+q^{17} \lambda ^2 (-2+\lambda +\lambda ^2)+q^{15} (\lambda +4 \lambda ^2-6 \lambda ^3)+q^{16} \lambda  (-1+\lambda +2 \lambda ^2-2 \lambda ^3)+q^2 (-2+3 \lambda +\lambda ^2-\lambda ^3)+q^3 (-1-6 \lambda +5 \lambda ^2+\lambda ^3)+q^4(5-2 \lambda -9 \lambda ^2+2 \lambda ^3)+q^{14} \lambda  (2-8 \lambda +2 \lambda ^2+5 \lambda ^3)+q^{11} \lambda  (6-19 \lambda -4 \lambda ^2+9 \lambda ^3)+q^7 (5-13 \lambda -12 \lambda ^2+12 \lambda ^3)+q^9 (-4+3 \lambda +25 \lambda ^2-9 \lambda ^3-5 \lambda ^4)+q^{13} (1-5 \lambda +3 \lambda ^2+10 \lambda ^3-5 \lambda ^4)+q^{12} (-2+3 \lambda +12 \lambda ^2-14 \lambda ^3-3 \lambda ^4)+q^{10} (3-12 \lambda +20 \lambda ^3-3 \lambda ^4)+q^6 (-5-5 \lambda +19 \lambda ^2+\lambda ^3-2 \lambda ^4)+q^5 (-2+13 \lambda -2 \lambda ^2-6 \lambda ^3+\lambda ^4)+q^8 (1+13 \lambda -17 \lambda ^2-12 \lambda ^3+5 \lambda ^4)\bigl]$
\end{tabular}  
\end{center}
\begin{center}
\begin{tabular}{c p{11cm}}
$V_{(\yng(2),\,\yng(2))}^{U(N)}[7_{3}]=$ &$\frac{(-1+\lambda ) (-1+q \lambda )}{(-1+q)^4 q^6 (1+q)^2 \lambda ^6}\bigl[(-1+\lambda ) \lambda -q^{12} \lambda ^3 (4-4 \lambda -3 \lambda ^2+\lambda ^3)+q (1+\lambda -3 \lambda ^2+\lambda ^3)+q^{13} \lambda ^3 (-1-3 \lambda +\lambda ^2+\lambda ^3)+q^2 (-2+4 \lambda -3 \lambda ^3+\lambda ^4)-q^{11} \lambda ^2 (2-8 \lambda -8 \lambda ^2+5 \lambda ^3+\lambda ^4)-q^3 (1+5 \lambda -9 \lambda ^2+3 \lambda ^4)+q^9 \lambda  (2+3 \lambda -20 \lambda ^2-4 \lambda ^3+7 \lambda ^4)+q^{14} (\lambda ^3-\lambda ^5)+q^4 (4-5 \lambda -8 \lambda ^2+11 \lambda ^3+\lambda ^4-\lambda ^5)-q^8 \lambda  (-2+13 \lambda -3 \lambda ^2-22 \lambda ^3+\lambda ^4+\lambda ^5)+q^5 (-1+9 \lambda -10 \lambda ^2-10 \lambda ^3+9 \lambda ^4+\lambda ^5)+q^6 (-2+\lambda +16 \lambda ^2-13 \lambda ^3-12 \lambda ^4+2 \lambda ^5)+q^{10} \lambda  (-1+4 \lambda +5 \lambda ^2-15 \lambda ^3-3 \lambda ^4+2 \lambda ^5)+q^7 (1-7 \lambda +3 \lambda ^2+22 \lambda ^3-7 \lambda ^4-5 \lambda ^5+\lambda ^6)\bigl]$
\end{tabular}  
\end{center}
\end{enumerate}
\renewcommand{\baselinestretch}{1}\selectfont
\Yboxdim5pt
Changing both the rank two symmetric representation $\yng(2)$ by 
antisymmetric representation $\yng(1,1)$, we find the following relation 
between the link polynomials:
\Yboxdim4pt
\begin{align}
\Yvcentermath1
V_{\left(\yng(2),\yng(2)\right)}^{U(N)}[\mathcal{L}](q,\lambda)=V_{\left(\yng(1,1),\yng(1,1)\right)}^{U(N)}[\mathcal{L}](q^{-1},\lambda).
\end{align} 

With these polynomial invariants available for the non-torus knots and links in Figures \ref{fig:knots}, \ref{fig:links},
we are in a position to verify Ooguri-Vafa\cite{Ooguri:1999bv} and Labastida-Marino-Vafa 
\cite{Labastida:2000yw} conjectures.

\section{Reformulated link invariants}
\label{sec:5}
\Yboxdim5pt
In this section we explcitly write  the reformulated link invariant for the  non torus
knots and links in Figure \ref{fig:knots} and Figure \ref{fig:links}.
Rewriting the most general form of reformulated invariants $f_R[\mathcal{K}]$   and $f_{R_1,R_2}[{\cal L}]$ (see equation (3.16) in \cite{Labastida:2001ts}) 
for representations $\yng(1)$ and $\yng(2)$ on the component
knots, the expression  for knots are
\Yboxdim4pt
\begin{eqnarray}
f_{\yng(1)}[\mathcal{K}] & = & V_{\yng(1)}[\mathcal{K}]\label{eqn:fsingle}\\
f_{\yng(2)}[\mathcal{K}] & = & V_{\yng(2)}[\mathcal{K}]-\frac{1}{2}\left(V_{\yng(1)}[\mathcal{K}]^{2}+V_{\yng(1)}^{(2)}[\mathcal{K}]\right)\\
f_{\yng(1,1)}[\mathcal{K}] & = & V_{\yng(1,1)}[\mathcal{K}]-\frac{1}{2}\left(V_{\yng(1)}[\mathcal{K}]^{2}-V_{\yng(1)}^{(2)}[\mathcal{K}]\right),
\end{eqnarray}
where we have suppressed $U(N)$ superscript on the knot invariants $(V_R[\mathcal{K}]\equiv
V_R^{U(N)}[\mathcal{K}](q,\lambda))$. Further,  we use the notation $V_{R}^{(n)}[\mathcal{K}]\equiv V_{R}[\mathcal{K}](q^n,\lambda^n)$. 
Ooguri-Vafa conjecture \cite{Ooguri:1999bv} states  that the  reformulated invariants for knots should have the following structure

\begin{align}
f_{R}(q,\lambda)=\sum_{s,Q}\frac{N_{Q,R,s}}{q^{1/2}-q^{-1/2}}a^{Q}q^{s},
\label{OVconj}
\end{align}
where $N_{Q,R,s}$ are integer and $Q$ and $s$ are, in general,
half integers. Clearly for \Yboxdim5pt$R=\yng(1)$ (fundamental representation), the polynomial structure in section \ref{sec:3} for 
$V_{\yng(1)}^{U(N)}[\mathcal{K}]$ (\ref{eqn:fsingle}) obeys eqn.(\ref{OVconj}). We will verify for \Yboxdim5pt$R=\yng(2)$ and $\yng(1,1)$ in 
the following subsection. 

The reformulated invariants in terms of two-component link invariants 
has the following form for $R_1,R_2 \in \{\yng(1),\yng(2), \yng(1,1)\}$:
\Yboxdim4pt
\begin{eqnarray}
f_{(\yng(1),\,\yng(1))}[\mathcal{L}] & = & V_{(\yng(1),\yng(1))}[\mathcal{L}]-V_{\yng(1)}[\mathcal{K}_{1}]V_{\yng(1)}[\mathcal{K}_{2}]\\
f_{(\yng(1),\,\yng(2))}[\mathcal{L}] & = & V_{(\yng(1),\yng(2))}[\mathcal{L}]-V_{(\yng(1),\yng(1))}[\mathcal{L}]V_{\yng(1)}[\mathcal{K}_{2}]-V_{\yng(1)}[\mathcal{K}_{1}]V_{\yng(2)}[\mathcal{K}_{2}]\nonumber \\
 &  & +V_{\yng(1)}[\mathcal{K}_{1}]V_{\yng(1)}[\mathcal{K}_{2}]^{2}\\
f_{\left(\yng(1),\,\yng(1,1)\right)}[\mathcal{L}] & = & V_{\left(\yng(1),\yng(1,1)\right)}[\mathcal{L}]-V_{(\yng(1),\yng(1))}[\mathcal{L}]V_{\yng(1)}[\mathcal{K}_{2}]-V_{\yng(1)}[\mathcal{K}_{1}]V_{\yng(1,1)}[\mathcal{K}_{2}]\nonumber \\
 &  & +V_{\yng(1)}[\mathcal{K}_{1}]V_{\yng(1)}[\mathcal{K}_{2}]^{2}\\
f_{(\yng(2),\,\yng(1))}[\mathcal{L}] & = & V_{(\yng(2),\yng(1))}[\mathcal{L}]-V_{(\yng(1),\yng(1))}[\mathcal{L}]V_{\yng(1)}[\mathcal{K}_{1}]-V_{\yng(2)}[\mathcal{K}_{1}]V_{\yng(1)}[\mathcal{K}_{2}]\nonumber \\
 &  & +V_{\yng(1)}[\mathcal{K}_{1}]^{2}V_{\yng(1)}[\mathcal{K}_{2}]\\
f_{\left(\yng(1,1),\,\yng(1)\right)}[\mathcal{L}] & = & V_{\left(\yng(1,1),\yng(1)\right)}[\mathcal{L}]-V_{(\yng(1),\yng(1))}[\mathcal{L}]V_{\yng(1)}[\mathcal{K}_{1}]-V_{\yng(1,1)}[\mathcal{K}_{1}]V_{\yng(1)}[\mathcal{K}_{2}]\nonumber \\
 &  & +V_{\yng(1)}[\mathcal{K}_{1}]^{2}V_{\yng(1)}[\mathcal{K}_{2}]
\end{eqnarray}
\begin{eqnarray}
f_{(\yng(2),\,\yng(2))}[\mathcal{L}] & = & V_{(\yng(2),\yng(2))}[\mathcal{L}]-V_{\yng(2)}(\mathcal{K}_{1})V_{\yng(2)}[\mathcal{K}_{2}]-V_{(\yng(2),\yng(1))}[\mathcal{L}]V_{\yng(1)}[\mathcal{K}_{2}]\nonumber \\
 &  & -V_{(\yng(1),\yng(2))}[\mathcal{L}]V_{\yng(1)}[\mathcal{K}_{1}]-\frac{1}{2}V_{(\yng(1),\yng(1))}[\mathcal{L}]^{2}+2V_{(\yng(1),\yng(1))}[\mathcal{L}]V_{\yng(1)}[\mathcal{K}_{1}]V_{\yng(1)}[\mathcal{K}_{2}]\nonumber \\
 &  & +V_{\yng(1)}[\mathcal{K}_{1}]^{2}V_{\yng(2)}[\mathcal{K}_{2}]+V_{\yng(2)}[\mathcal{K}_{1}]V_{\yng(1)}[\mathcal{K}_{2}]^{2}-\frac{3}{2}V_{(\yng(1))}[\mathcal{K}_{1}]^{2}V_{(\yng(1))}[\mathcal{K}_{2}]^{2}\nonumber \\
 &  & -\frac{1}{2}V_{(\yng(1),\yng(1))}^{(2)}[\mathcal{L}]+\frac{1}{2}V_{\yng(1)}^{(2)}[\mathcal{K}_{1}]V_{\yng(1)}^{(2)}[\mathcal{K}_{2}]
\end{eqnarray}
\begin{eqnarray}
f_{\left(\yng(1,1),\,\yng(1,1)\right)}[\mathcal{L}] & = & V_{\left(\yng(1,1),\yng(1,1)\right)}[\mathcal{L}]-V_{\yng(1,1)}[\mathcal{K}_{1}]V_{\yng(1,1)}[\mathcal{K}_{2}]-V_{\left(\yng(1,1),\yng(1)\right)}[\mathcal{L}]V_{\yng(1)}[\mathcal{K}_{2}]\nonumber \\
 &  & -V_{\left(\yng(1),\yng(1,1)\right)}[\mathcal{L}]V_{\yng(1)}[\mathcal{K}_{1}]-\frac{1}{2}V_{(\yng(1),\yng(1))}[\mathcal{L}]^{2}+2V_{(\yng(1),\yng(1))}[\mathcal{L}]V_{\yng(1)}[\mathcal{K}_{1}]V_{\yng(1)}[\mathcal{K}_{2}]\nonumber \\
 &  & +V_{\yng(1)}[\mathcal{K}_{1}]^{2}V_{\yng(1,1)}[\mathcal{K}_{2}]+V_{\yng(1,1)}[\mathcal{K}_{1}]V_{\yng(1)}[\mathcal{K}_{2}]^{2}-\frac{3}{2}V_{(\yng(1))}[\mathcal{K}_{1}]^{2}V_{(\yng(1))}[\mathcal{K}_{2}]^{2}\nonumber \\
 &  & -\frac{1}{2}V_{(\yng(1),\yng(1))}^{(2)}[\mathcal{L}]+\frac{1}{2}V_{\yng(1)}^{(2)}[\mathcal{K}_{1}]V_{\yng(1)}^{(2)}[\mathcal{K}_{2}]
\end{eqnarray}
Here the components knots $\mathcal{K}_1$ and $\mathcal{K}_2$ are unknots for the non-torus links in Figure \ref{fig:links}.
The generalisation of Ooguri-Vafa  conjecture for links was proposed in \cite{Labastida:2000yw}  which 
states that reformulated invariants for  $r$-component link should have the following structure
  \begin{align}
f_{(R_1, R_2, \ldots R_r)} (q, \lambda)= (q^{1/2}- q^{-1/2})^{r-2}
\sum_{Q,s} N_{(R_1, \ldots R_r),Q,s} q^s \lambda^Q~, \label{reformulated}
\end{align}
where $N_{(R_1,\cdots R_r),Q,s}$ are integer and $Q$ and $s$ are half integers.
We can see below that all the reformulated invariants we calculate indeed satisfy the conjecture.
\subsection{Reformulated invariant for knots }

We have already seen in section \ref{sec:3}, $V_{\yng(1)}[\mathcal{K}]$ has the Ooguri-Vafa form given in eqn.(\ref{OVconj}). 
\Yboxdim5pt
For the symmetric second rank tensor $R=\yng(2)$ placed on the knot, \Yboxdim4pt$f_{\yng(2)}[\mathcal{K}]$ are: 
\renewcommand{\baselinestretch}{1.5}\selectfont
\Yboxdim4pt
\begin{center}
\begin{tabular}{c p{11cm}}
$f_{\yng(2)}[4_{1}]=$ &$\frac{(-1+\lambda )^2 }{(-1+q) q^2 \lambda ^3}\left[-q+\lambda -q^5 \lambda ^3+q^4 \lambda ^4+q^3 \lambda  (1+\lambda )-q^2 \lambda ^2 (1+\lambda )\right]$
\end{tabular}  
\end{center}
\begin{center}
\begin{tabular}{c p{11cm}}
$f_{\yng(2)}[5_{2}]=$ &$-\frac{(1-q+q^2) (-1+\lambda )^2 }{(-1+q) q^5 \lambda ^7}\bigl[q (-1+\lambda )+q^2 (-1+\lambda )+\lambda +q^4 \lambda (1+\lambda +\lambda ^2)-q^3 (1+\lambda ^2+\lambda ^3+\lambda ^4)\bigl]$
\end{tabular}  
\end{center}
\begin{center}
\begin{tabular}{c p{11cm}}
$f_{\yng(2)}[6_{1}]=$ &$-\frac{(-1+\lambda )^2 }{(-1+q) q^4 \lambda ^5}\bigl[q-\lambda -q^2 \lambda +q^7 \lambda ^4 (1+\lambda )+q^3 (1+\lambda ^2)-q^5 \lambda  (1+\lambda +\lambda ^2)^2+q^4 \lambda(-1+\lambda +2 \lambda ^2+2 \lambda ^3+\lambda ^4)+q^6 (\lambda ^2+\lambda ^3+\lambda ^4-\lambda ^5-\lambda ^6\bigl]$
\end{tabular}  
\end{center}
\begin{center}
\begin{tabular}{c p{11cm}}
$f_{\yng(2)}[6_{2}]=$ &$-\frac{(1-q+q^2) (-1+\lambda )^2 }{(-1+q) q^6 \lambda ^5}\bigl[-\lambda -q^5 \lambda +q^2 (-1+\lambda ) \lambda +q^9 \lambda ^3-q^8 (-1+\lambda ) \lambda ^3+q^6 \lambda ^2 (1+\lambda )+q (1+\lambda ^2)-q^3 \lambda  (2+\lambda ^2)+q^4 (1+2 \lambda ^2)-q^7 \lambda  (1+\lambda +\lambda ^3)\bigl]$  
\end{tabular}
\end{center}
\begin{center}
\begin{tabular}{c p{11cm}}
$f_{\yng(2)}[6_{3}]=$ &$-\frac{(1-q+q^2) (-1+\lambda )^2 }{(-1+q) q^5 \lambda ^3}\bigl[-\lambda +q^2 \lambda ^2-q^7 \lambda ^2+q^9 \lambda ^3+q (1+\lambda ^2)+q^3 (-1-\lambda +\lambda ^2)+q^6 \lambda ^2 (-1+\lambda +\lambda ^2)-q^4 \lambda  (2+2 \lambda +2 \lambda ^2+\lambda ^3)+q^5 (1+2 \lambda +2 \lambda ^2+2 \lambda ^3)-q^8 (\lambda ^2+\lambda ^4)\bigl]$  
\end{tabular}
\end{center}
\begin{center}
\begin{tabular}{c p{11cm}}
$f_{\yng(2)}[7_{2}]=$ &$-\frac{(-1+\lambda )^2 }{(-1+q) q^7 \lambda ^9}\bigl[q-\lambda -q^2 \lambda -2 q^4 \lambda +q^3 (1+\lambda ^2)-q^8 \lambda  (1+\lambda +2 \lambda ^2+2 \lambda ^3+\lambda ^4)-q^6 \lambda  (2+\lambda +3 \lambda ^2+3 \lambda ^3+3 \lambda ^4+\lambda ^5)+q^5 (1+\lambda ^2+\lambda ^4+\lambda ^5+\lambda ^6)+q^7 (1+\lambda +3 \lambda ^2+3 \lambda ^3+4 \lambda ^4+3 \lambda ^5+\lambda ^6)\bigl]$  
\end{tabular}
\end{center}
\begin{center}
\begin{tabular}{c p{11cm}}
$f_{\yng(2)}[7_{3}]=$ &$\frac{(-1+\lambda )^2 \lambda ^3 }{(-1+q) q^2}\bigl[-2 q^{10} \lambda ^3-q^{12} \lambda ^3+q^{11} \lambda ^4-\lambda  (1+\lambda +\lambda ^2)+q^6 \lambda  (-1-\lambda -5 \lambda ^2+\lambda ^3)+q^4 \lambda  (-1-\lambda -4 \lambda ^2+\lambda ^3)-2 q^2 (1+2 \lambda +2 \lambda ^2+2 \lambda ^3)+q^5 \lambda  (1+2 \lambda -\lambda ^2+3 \lambda ^3)+q^8 (-1-\lambda -\lambda ^2-3 \lambda ^3+\lambda ^4)+q (1+3 \lambda +3 \lambda ^2+\lambda ^3+\lambda ^4)+q^3 (2+3 \lambda +4 \lambda ^2+2 \lambda ^4)+q^9 (1+\lambda +\lambda ^2-\lambda ^3+2 \lambda ^4)+q^7 (1+2 \lambda +2 \lambda ^2-\lambda ^3+2 \lambda ^4)\bigl]$  
\end{tabular}
\end{center}
\begin{center}
\begin{tabular}{c p{11cm}}
$f_{\yng(2)}[7_{4}]=$ &$\frac{(-1+\lambda )^2 \lambda  }{(-1+q) q}\bigl[-q^9 \lambda ^5+\lambda ^2 (1+\lambda )-q^7 \lambda ^3 (1+\lambda )+q^8 \lambda ^4 (-1+\lambda ^2)+q^6 \lambda ^2(1+\lambda +\lambda ^2+\lambda ^3+\lambda ^4)-q \lambda (2+4 \lambda +4 \lambda ^2+3 \lambda ^3+\lambda ^4)+q^5(\lambda +\lambda ^2-2 \lambda ^4-\lambda ^5)-q^3 (2+5 \lambda +5 \lambda ^2+5 \lambda ^3+4 \lambda ^4+\lambda ^5)+q^4 (1+\lambda +2 \lambda ^2+\lambda ^3-\lambda ^4+\lambda ^5+\lambda ^6)+q^2(1+5 \lambda +8 \lambda ^2+7 \lambda ^3+3 \lambda ^4+2 \lambda ^5+\lambda ^6)\bigl]$  
\end{tabular}
\end{center}
\begin{center}
\begin{tabular}{c p{11cm}}
$f_{\yng(2)}[7_{5}]=$ &$\frac{(1-q+q^2) (-1+\lambda )^2 }{(-1+q) q^9 \lambda ^9}\bigl[-q+\lambda -q^7 (-1+\lambda ) \lambda ^3+q^8 \lambda  (1+\lambda )^3+q^{10} \lambda  (1+2 \lambda +2 \lambda ^2)+q^4 (-1+\lambda +\lambda ^2+2 \lambda ^3)+q^6 (\lambda ^3-2 \lambda ^4)+q^3 (\lambda -\lambda ^4)-q^5 (1+\lambda +\lambda ^2+\lambda ^4)-q^9 (1+2 \lambda +3 \lambda ^2+3 \lambda ^3+2 \lambda ^4)\bigl]$  
\end{tabular}
\end{center}
\begin{center}
\begin{tabular}{c p{11cm}}
$f_{\yng(2)}[7_{6}]=$ &$-\frac{(-1+\lambda )^2}{(-1+q) q^6 \lambda ^7}\bigl[-(-1+\lambda ) \lambda ^2+q^{11} \lambda ^5-q^{10} \lambda ^5 (1+\lambda )+q^9 \lambda ^3 (-2-2 \lambda -\lambda ^2+\lambda ^3)+q \lambda  (-2+\lambda +\lambda ^2+\lambda ^3)-q^3 \lambda (1-3 \lambda +\lambda ^2+\lambda ^4)-q^2(-1+\lambda ^2+3 \lambda ^3+\lambda ^4)+q^8 \lambda ^2 (2+4 \lambda +5 \lambda ^2+4 \lambda ^3+\lambda ^4)-q^5 \lambda (3+4 \lambda +8 \lambda ^2+6 \lambda ^3+2 \lambda ^4)-q^7 \lambda (2+2 \lambda +7 \lambda ^2+6 \lambda ^3+4 \lambda ^4+2 \lambda ^5)+q^4 (1-\lambda +4 \lambda ^2+3 \lambda ^3+5 \lambda ^4+2 \lambda ^5)+q^6 (1+\lambda +6 \lambda ^2+6 \lambda ^3+7 \lambda ^4+2 \lambda ^5+\lambda ^6)\bigl]$  
\end{tabular}
\end{center}
\begin{center}
\begin{tabular}{c p{11cm}}
$f_{\yng(2)}[7_{7}]=$ &$\frac{(-1+\lambda )^2 }{(-1+q) q^5 \lambda ^3}\bigl[\lambda +2 q^3 \lambda +q^{11} (-1+\lambda ) \lambda ^3-q (1+\lambda )^2+q^2 (2+\lambda)-q^8 \lambda ^2 (-1+\lambda -2 \lambda ^2+\lambda ^3)+q^9 \lambda ^2 (-1-2 \lambda -\lambda ^2-\lambda ^3+\lambda ^4)-q^5 \lambda  (7+12 \lambda +12 \lambda ^2+5 \lambda ^3+2 \lambda ^4)+q^4 (-2+5 \lambda ^2+5 \lambda ^3+4 \lambda ^4)+q^{10} (\lambda ^2+\lambda ^3+2 \lambda ^4-2 \lambda ^5)+q^6 (2+8 \lambda +11 \lambda ^2+9 \lambda ^3+4 \lambda ^4-\lambda ^5)+q^7 (-1-3 \lambda -4 \lambda ^2-3 \lambda ^3+3 \lambda ^4-\lambda ^5+\lambda ^6)\bigl]$  
\end{tabular}
\end{center}
\begin{center}
\begin{tabular}{c p{11cm}}
$f_{\yng(2)}[8_{1}]=$ &$-\frac{(-1+\lambda )^2 }{(-1+q) q^6 \lambda ^7}\bigl[q-\lambda -q^2 \lambda -2 q^4 \lambda +q^3 (1+\lambda ^2)+q^5 (1+\lambda ^2)+q^9 \lambda ^5 (1+\lambda +\lambda ^2)-q^8 \lambda ^2 (-1-2 \lambda -2 \lambda ^2-2 \lambda ^3+\lambda ^5+\lambda ^6)+q^6 \lambda  (-1+\lambda +2 \lambda ^2+3 \lambda ^3+3 \lambda ^4+2 \lambda ^5+\lambda ^6)-q^7 \lambda  (1+2 \lambda +4 \lambda ^2+5 \lambda ^3+4 \lambda ^4+3 \lambda ^5+\lambda ^6)\bigl]$  
\end{tabular}
\end{center}
\begin{center}
\begin{tabular}{c p{11cm}}
$f_{\yng(2)}[9_{2}]=$ &$-\frac{(-1+\lambda )^2}{(-1+q) q^9 \lambda ^{11}} \bigl[q-\lambda -q^2 \lambda -2 q^4 \lambda -2 q^6 \lambda +q^3 (1+\lambda ^2)+q^5(1+\lambda ^2)-q^{10} \lambda  (1+\lambda +2 \lambda ^2+3 \lambda ^3+3 \lambda ^4+2 \lambda ^5+\lambda ^6)-q^8 \lambda  (2+\lambda +2 \lambda ^2+4 \lambda ^3+4 \lambda ^4+4 \lambda ^5+3 \lambda ^6+\lambda ^7)+q^7 (1+\lambda ^2+\lambda ^5+\lambda ^6+\lambda ^7+\lambda ^8)+q^9(1+\lambda +3 \lambda ^2+4 \lambda ^3+5 \lambda ^4+6 \lambda ^5+5 \lambda ^6+3 \lambda ^7+\lambda ^8)\bigl]$  
\end{tabular}
\end{center}
\begin{center}
\begin{tabular}{c p{11cm}}
$f_{\yng(2)}[10_{1}]=$ &$-\frac{(-1+\lambda )^2}{(-1+q) q^8 \lambda ^9} \bigl[q-\lambda -q^2 \lambda -2 q^4 \lambda -2 q^6 \lambda +q^3 (1+\lambda ^2)+q^5 (1+\lambda ^2)+q^7 (1+\lambda ^2)+q^{11} \lambda ^6 (1+\lambda +\lambda ^2+\lambda ^3)+q^{10} \lambda ^2 (1+2 \lambda +3 \lambda ^2+3 \lambda ^3+3 \lambda ^4+\lambda ^5-\lambda ^7-\lambda ^8)+q^8 \lambda  (-1+\lambda +2 \lambda ^2+3 \lambda ^3+4 \lambda ^4+4 \lambda ^5+3 \lambda ^6+2 \lambda ^7+\lambda ^8)-q^9 \lambda  (1+2 \lambda +4 \lambda ^2+6 \lambda ^3+7 \lambda ^4+6 \lambda ^5+5 \lambda ^6+3 \lambda ^7+\lambda ^8\bigl]$  
\end{tabular}
\end{center}

\renewcommand{\baselinestretch}{1}\selectfont
\Yboxdim5pt
Changing the symmetric representation by antisymmetry representation, we find  the
following relation between the reformulated invariants:

\begin{align}
\Yboxdim4pt
f_{\yng(2)}[\mathcal{K}](q^{-1},\lambda)=f_{\yng(1,1)}[\mathcal{K}](q,\lambda).
\end{align}

\subsection{Reformulated invariant for links}

\renewcommand{\baselinestretch}{1.5}\selectfont
\begin{enumerate}

\Yboxdim6pt
\item
For $R_1=\yng(1)$, $R_2=\yng(1)$    : 
\Yboxdim4pt

\begin{center}
\begin{tabular}{c p{11cm}}
$f_{(\yng(1),\yng(1))}[6_{2}]=$ &$\frac{(-1+\lambda ) }{q \lambda }\bigl[-q+\lambda +q^2 \lambda +\lambda ^2+q^2 \lambda ^2\bigl]$  
\end{tabular}
\end{center}

\begin{center}
\begin{tabular}{c p{11cm}}
$f_{(\yng(1),\yng(1))}[6_{3}]=$ &$\frac{(-1+\lambda ) }{q \lambda ^3}\bigl[\lambda +q^2 \lambda -q \left(1+\lambda +2 \lambda ^2\right)\bigl]$  
\end{tabular}
\end{center}

\begin{center}
\begin{tabular}{c p{11cm}}
$f_{(\yng(1),\yng(1))}[7_{1}]=$ &$\frac{(-1+\lambda ) }{q^2 \lambda ^2}\bigl[-\lambda -q^4 \lambda +q^2 (-2+\lambda ) \lambda +q (1+\lambda ^2)+q^3 (1+\lambda ^2)\bigl]$  
\end{tabular}
\end{center}

\begin{center}
\begin{tabular}{c p{11cm}}
$f_{(\yng(1),\yng(1))}[7_{2}]=$ &$\frac{(-1+\lambda ) }{q^2 \lambda ^2}\bigl[(-\lambda -3 q^2 \lambda -q^4 \lambda +q (1+\lambda ^2)+q^3(1+\lambda ^2)\bigl]$  
\end{tabular}
\end{center}

\begin{center}
\begin{tabular}{c p{11cm}}
$f_{(\yng(1),\yng(1))}[7_{3}]=$ &$\frac{\left(-1+\lambda ^2\right) }{q \lambda ^3}\bigl[q-\lambda -q^2 \lambda +q \lambda ^2\bigl]$  
\end{tabular}
\end{center}

\Yboxdim6pt
\item
For $R_1=\yng(1)$, $R_2=\yng(2)$    : 
\Yboxdim4pt

\begin{center}
\begin{tabular}{c p{11cm}}
$f_{(\yng(1),\yng(2))}[6_{2}]=$ &$ \frac{(-1+\lambda ) }{\sqrt{q} \sqrt{\lambda }}\bigl[\lambda ^2+q \lambda ^2+q^3 \lambda ^2+q^4 \lambda ^2+q^2 \left(-1-\lambda +\lambda ^2\right)\bigl]$
\end{tabular}
\end{center}

\begin{center}
\begin{tabular}{c p{11cm}}
$f_{(\yng(1),\yng(2))}[6_{3}]=$ &$\frac{1}{q^{5/2} \lambda ^{7/2}}(1+q) (-1+\lambda ) \bigl[\lambda +q^2 \lambda -q \left(1+\lambda +\lambda ^2\right)\bigl]$  
\end{tabular}
\end{center}

\begin{center}
\begin{tabular}{c p{11cm}}
$f_{(\yng(1),\yng(2))}[7_{1}]=$ &$\frac{(-1+\lambda ) }{q^{5/2} \lambda ^{5/2}}\bigl[q-\lambda -q^6 \lambda ^2+q^3 (-2+\lambda ) \lambda ^2+q^2 \lambda ^3+q^4 \lambda  (1-\lambda +\lambda ^2)+q^5 \lambda  (1-\lambda +\lambda ^2)\bigl]$  
\end{tabular}
\end{center}

\begin{center}
\begin{tabular}{c p{11cm}}
$f_{(\yng(1),\yng(2))}[7_{2}]=$ &$\frac{(-1+\lambda )}{q^{7/2} \lambda ^{5/2}} \bigl[(q^4-\lambda -3 q^3 \lambda -q^6 \lambda ^2+q^5 \lambda ^3+q (1-\lambda +\lambda ^2)+q^2 (1-\lambda +\lambda ^2)\bigl]$  
\end{tabular}
\end{center}

\begin{center}
\begin{tabular}{c p{11cm}}
$f_{(\yng(1),\yng(2))}[7_{3}]=$ &$\frac{\left(-1+\lambda ^2\right) }{q^{3/2} \lambda ^{7/2}}\bigl[q-\lambda -q^3 \lambda ^2+q^2 \lambda ^3\bigl]$  
\end{tabular}
\end{center}

\renewcommand{\baselinestretch}{1}\selectfont
We have checked that $f_{(\yng(1),\yng(2))}[L]=f_{(\yng(2),\yng(1))}[L]$ for these links. We also have the symmetry relation 
\begin{align}
\Yvcentermath1
f_{(\yng(1),\yng(2))}[\mathcal{L}](q^{-1},\lambda)=-f_{\left(\yng(1),\yng(1,1)\right)}[\mathcal{L}](q,\lambda).
\end{align}

\Yboxdim6pt
\item
For $R_1=\yng(2)$, $R_2=\yng(2)$    : 
\Yboxdim4pt

\begin{center}
\begin{tabular}{c p{11cm}}
$f_{(\yng(2),\yng(2))}[6_{2}]=$ &$\frac{1}{q^2 \lambda }\bigl[q (-1+\lambda )^2 \lambda +q^9 (-1+\lambda )^2 \lambda ^2+\lambda ^3-\lambda ^5+q^7 (-1+\lambda )^2 \lambda ^2 (3+\lambda )+q^{10} \lambda ^3 (-1+\lambda ^2)+q^8 \lambda ^2 (2-3 \lambda +\lambda ^2)+q^5 (-1+\lambda )^2 (-1-2 \lambda +2 \lambda ^2)+q^3 (-1+\lambda )^2 (1+\lambda +2 \lambda ^2)+q^4 (-1+\lambda +4 \lambda ^2-5 \lambda ^3+\lambda ^4)+q^6 \lambda  (-1+5 \lambda -8 \lambda ^2+3 \lambda ^3+\lambda ^4)+q^2 (-1+\lambda ^2+\lambda ^4-\lambda ^5)\bigl]$
\end{tabular}
\end{center}

\begin{center}
\begin{tabular}{c p{11cm}}
$f_{(\yng(2),\yng(2))}[6_{3}]=$ &$\frac{1}{q^7 \lambda ^6}\bigl[-(-1+\lambda )^2 \lambda +q^2 (-1+\lambda )^2 \lambda ^2-q^9 (-1+\lambda ) \lambda ^3+q^5 (-1+\lambda )^2 \lambda ^2 (2+\lambda ^2)-q^4 (-1+\lambda )^2 \lambda  (1-\lambda +2 \lambda ^2)+q^8 (-1+\lambda )^2 \lambda (1+\lambda +2 \lambda ^2)+q^3 \lambda ^2 (3-4 \lambda +3 \lambda ^2-2 \lambda ^3)+q^6 (-1+\lambda )^2 \lambda  (1+3 \lambda +2 \lambda ^2+2 \lambda ^3)+q (1-2 \lambda +\lambda ^2-\lambda ^3+\lambda ^4)-q^7 (1+2 \lambda ^2-2 \lambda ^3-2 \lambda ^4+\lambda ^6)\bigl]$  
\end{tabular}
\end{center}

\begin{center}
\begin{tabular}{c p{11cm}}
$f_{(\yng(2),\yng(2))}[7_{1}]=$ &$\frac{1}{q^6 \lambda ^4}\bigl[-(-1+\lambda )^2 \lambda +q^{11} (-1+\lambda )^4 \lambda ^2-q^{13} (-1+\lambda )^2 \lambda ^3+q (-1+\lambda )^2 (1+\lambda ^2)+3 q^9 (-1+\lambda )^2 \lambda ^2(1-\lambda +\lambda ^2)+q^3 (-1+\lambda )^2 (1-\lambda +2 \lambda ^2)+q^2 \lambda  (-2+4 \lambda -3 \lambda ^2+\lambda ^3)+q^{12} \lambda ^2 (1-2 \lambda +3 \lambda ^2-3 \lambda ^3+\lambda ^4)-q^6 (1+\lambda -\lambda ^2-2 \lambda ^3+\lambda ^4)+q^7 (-1+\lambda )^2 (1+2 \lambda +2 \lambda ^2-\lambda ^3+\lambda ^4)+q^8 \lambda ^2 (2-4 \lambda +7 \lambda ^2-7 \lambda ^3+2 \lambda ^4)-q^5 \lambda (-1+\lambda +2 \lambda ^3-3 \lambda ^4+\lambda ^5)+q^{10} \lambda  (-1+2 \lambda -7 \lambda ^2+11 \lambda ^3-7 \lambda ^4+2 \lambda ^5)-q^4 (-1+3 \lambda -4 \lambda ^2+\lambda ^3+\lambda ^4-\lambda ^5+\lambda ^6)\bigl]$  
\end{tabular}
\end{center}

\begin{center}
\begin{tabular}{c p{11cm}}
$f_{(\yng(2),\yng(2))}[7_{2}]=$ &$-\frac{1}{q^8 \lambda ^4}\bigl[(-1+\lambda )^2 \lambda +q^{13} (-1+\lambda )^2 \lambda ^3-q (-1+\lambda )^2 (1+\lambda ^2)+q^3 (-1+\lambda )^2 \lambda  (3-2 \lambda +\lambda ^2)+q^{11} (-1+\lambda )^2 \lambda  (-1-\lambda +\lambda ^2)+q^5 (-1+\lambda )^2 (-1+2 \lambda -5 \lambda ^2+\lambda ^3)-q^2 \lambda  (-1+4 \lambda -4 \lambda ^2+\lambda ^3)-q^{12} \lambda ^3 (-1+2 \lambda -2 \lambda ^2+\lambda ^3)-q^7 (-1+\lambda )^2 (1+\lambda +5 \lambda ^2+\lambda ^3)+q^{10} \lambda ^2 (1-2 \lambda -2 \lambda ^2+3 \lambda ^3)+q^6 \lambda  (3-10 \lambda +9 \lambda ^2-5 \lambda ^3+3 \lambda ^4)+q^4 (-1+6 \lambda -11 \lambda ^2+12 \lambda ^3-7 \lambda ^4+\lambda ^5)-q^9 \lambda  (3-3 \lambda -2 \lambda ^3+\lambda ^4+\lambda ^5)+q^8 (1+\lambda +2 \lambda ^2-5 \lambda ^3+\lambda ^4-\lambda ^5+\lambda ^6)\bigl]$  
\end{tabular}
\end{center}

\begin{center}
\begin{tabular}{c p{11cm}}
$f_{(\yng(2),\yng(2))}[7_{3}]=$ &$-\frac{(-1+\lambda )^2 }{q^6 \lambda ^6}\bigl[(-q+\lambda -q^5 \lambda +q^9 \lambda ^4 (1+\lambda )+q^4 \lambda  (1+\lambda )^2+q^6 \lambda ^2 (1+\lambda )^2 (1+\lambda ^2)-q^3 \lambda ^2 (1+\lambda +\lambda ^2)-q^7 \lambda  (1+2 \lambda +3 \lambda ^2+3 \lambda ^3+3 \lambda ^4+\lambda ^5)+q^8 (\lambda ^2+\lambda ^3+2 \lambda ^4-\lambda ^6)\bigl]$  
\end{tabular}
\end{center}

\renewcommand{\baselinestretch}{1}\selectfont
\Yboxdim5pt
Changing both the rank two symmetric representation $\yng(2)$ by 
antisymmetric representation $\yng(1,1)$, we find  the
following relation between the reformulated invariants:
\Yboxdim4pt
\begin{align}
\Yvcentermath1
f_{\left(\yng(2),\yng(2)\right)}[\mathcal{L}](q,\lambda)=f_{\left(\yng(1,1),\yng(1,1)\right)}[\mathcal{L}](q^{-1},\lambda).
\end{align} 

\end{enumerate}

\renewcommand{\baselinestretch}{1}\selectfont
\section{Conclusion}
\Yboxdim5pt
This paper presents the $U(N)$ Chern-Simons invariants for the non-torus knots and non-torus links in Figure \ref{fig:knots} and Figure \ref{fig:links}.  We have written the explicit polynomial form for few representations $(\one,\twohor,\twover)$  and also obtained the reformulated invariants. For completeness, we have included reformulated invariants of knots $4_1,6_1$ \cite{Ramadevi:2000gq}, knot $5_2$ and link $6_1$  \cite{Zodinmawia:2011ud}.   The form of these reformulated invariants  are consistent with the
 Ooguri-Vafa (\ref{OVconj})  and Labastida-Marino-Vafa conjectures (\ref{reformulated}). 

Recently, the polynomial form for figure eight knot carrying totally symmetric representation is given in Ref.\cite{Itoyama:2012fq}. Further,  the polynomial 
for the  knot carrying totally antisymmetric representation can be obtained 
by suitable change of polynomial variable which we also observe when we change $R=\twohor$ to $R=\twover$(\ref{symm}).
Motivated by the conjecture of the polynomial form for figure eight,  we have been trying to propose  the polynomial forms for non-torus knots carrying $n$th-rank symmetric tensor representation. Particularly  looking at the pattern of the polynomials, for a class of twist knots $\mathcal {K}_p$  carrying  representations $R=\yng(1),\yng(2)$ and $\yng(1,1)$ , 
we attempted to write $V_n^{U(N)}[\mathcal {K}_p]$ \cite{Satoshi:2012} where subscript $n$ denotes the $n$-rank symmetric representation.
We leave the reader to see the forthcoming paper \cite{Satoshi:2012,Gukov:toappear} for interesting results on twist knots.  

There are other non-torus knots besides the twist knots in  Figure \ref{fig:knots}. We hope that the results  in Ref.\cite{Satoshi:2012}, will  
suggest a closed form expression for $SU(N)$ quantum Racah coefficients  similar to $SU(2)$  quantum Racah coefficients \cite{Kirillov:1989}. In fact, the closed form expression 
for $SU(N)$ quantum Racah coefficients will determine the colored  polynomial invariant for any knot  or any  link carrying arbitrary representation $R$.

In the light of the recent developments on superpolynomials \cite{Itoyama:2012fq,Fuji:2012pm,Satoshi:2012}, which has an additional  polynomial variable $t$,  it should be a systematic exercise to look at the $t$-deformation of the quantum Racah coefficients.  We hope to report on these aspects in future.

\bibliography{CS}{}
\bibliographystyle{JHEP}
\end{document}